\documentclass[aps,pra,twocolumn,amsmath,amssymb,superscriptaddress,showpacs]{revtex4-1}

\usepackage[utf8]{inputenc}
\usepackage{amsfonts}
\usepackage{amsmath}
\usepackage{amssymb}
\usepackage{amsthm}
\usepackage{fontenc}
\usepackage{graphicx}
\usepackage{xcolor}
\usepackage{textcomp}
\usepackage{epstopdf}
\usepackage{braket}
\usepackage{mathtools}
\usepackage{amsmath}
\usepackage{dcolumn}
\usepackage{multirow}

\begin{document}
\newcommand{\abs}[1]{\lvert#1\rvert}
\title{Defect-enhanced Rashba spin-polarized currents in carbon nanotubes}

\author{Hern\'an Santos}
\affiliation{Dep. de F\'{\i}sica Fundamental, Universidad Nacional de Educaci\' on a Distancia, Apartado 60141, E-28040 Madrid, Spain}
\affiliation{Instituto de F\' isica, Universidade Federal Fluminense, Niter\' oi, Av.~Litor\^anea sn 24210-340, RJ-Brazil}
\author{Leonor Chico}
\affiliation{Instituto de Ciencia de Materiales de Madrid, Consejo Superior de Investigaciones Cient\'{\i}ficas, C/ Sor Juana In\'es de la Cruz 3, 28049 Madrid, Spain}
\author{J. E. Alvarellos}
\affiliation{Dep. de F\'{\i}sica Fundamental, Universidad Nacional de Educaci\' on a Distancia, Apartado 60141, E-28040 Madrid, Spain}
\author{A. Latg\' e}
\email{andrea.latge@gmail.com}
\affiliation{Instituto de F\' isica, Universidade Federal Fluminense, Niter\' oi, Av.~Litor\^anea sn 24210-340, RJ-Brazil}

\date{\today}

\begin{abstract}

The production of spin-polarized currents in 
pristine carbon nanotubes with Rashba spin-orbit interaction has been shown to be very sensitive to the symmetry of the tubes and the geometry of the setup. Here we analyze 
the role of defects on the spin quantum conductances of metallic carbon nanotubes due to an external electric field. We show that localized defects, such as 
adsorbed hydrogen atoms or pentagon-heptagon pairs, increase the Rashba spin-polarized current. Moreover, this enhancement takes place for energies closer to the Fermi energy as compared to the 
response of pristine tubes.  
Such increment can be even larger when several equally spaced defects are 
introduced into the system. 
We explore different arrangements of defects, showing that for certain geometries there are 
 flips of the spin-polarized current and even transport suppression. Our results indicate that spin valve devices at the nanoscale may be achieved via defect engineering in carbon nanotubes. 
 
\end{abstract}

\maketitle

\section{Introduction}

The departure from ideal crystalline order in carbon materials 
with $sp^2$ hybridization, 
due to the presence of topological defects, lattice distortions and atomic 
adsorption 
have strong consequences in the physical properties of such nanosystems \cite{Padilha2011}. In particular, these defects can create magnetic moments, as it has been observed in graphene by means of spin currents \cite{Kathleen2012} or scanning tunneling microscopes (STM) \cite{Ugeda2010}.
Novel control and manipulation features of these microscopes \cite{Ugeda2010,Gonzalez2016} and advances in material growth techniques \cite{SanchezValencia2014,Ruffieux2016} are the basis of outstanding experimental results. 
For instance, in a recent experiment  the position of hydrogen adatoms in graphene has been modified with an STM, obtaining 
configurations in which all atoms were in a specific sublattice \cite{Gonzalez2016}. 
Topological defects in graphene have been also created by electron beam focusing, achieving different structural defects by tuning the exposure time \cite{Robertson2012}.
 Another experimental goal that is being recently attained is the ability to choose the particular geometry of nanoscale carbon materials. 
Typical examples are the chirality and radius of carbon nanotubes (CNTs)
\cite{SanchezValencia2014}, 
or the width and edge shape of graphene nanoribbons \cite{Ruffieux2016}.

Spin-polarized currents created by the applied electric field that  induce a Rashba spin-orbit (RSO) interaction in carbon materials have been recently predicted \cite{Chico2015,HSantos2016}, even in the absence of external magnetic fields or magnetic impurities. These results indicate the possibility of producing stable spin-polarized electrical currents from unpolarized charge currents, paving the way toward the creation of carbon-based spin valve devices at the nanoscale.  Such studies focused on ideal pristine carbon nanotubes and nanoribbons, for which the spin-polarized currents appear far from the Fermi energy ($E_F$). However, the role of 
defects in carbon materials under a RSO interaction has not yet been addressed; defects might produce spin polarization of the conductance at energies around $E_F$.

Vacancies, adsorption of hydrogen atoms, and defects in general act as localized perturbations in graphene systems and can induce localized states 
close to the Dirac point. The effects of such perturbations on the electronic properties of the system are greatly dependent on their position in the two sublattices of graphene and carbon nanotubes 
\cite{Lieb1989,Saremi2007,Palacios2008,HSantos2014}. 
When a $p_z$ orbital is removed from one sublattice of pristine graphene, due for instance to the creation of a vacancy, a new state appears on the other sublattice, localized around the perturbation 
\cite{Liang2012}. Hydrogen adatoms also produce such localization effects; indeed, if we are not interested in the analysis of the resulting magnetic moments, we may consider 
both situations as equivalent.

The aim of the present work is to investigate how CNTs with defects may produce spin-polarized currents due to RSO coupling, thus possibly opening up novel routes for engineering the spin-dependent transport response on CNTs. We explore the possibility of enhancing such spin-polarized currents in CNTs closer to $E_F$  by introducing topological defects or hydrogen adatoms (H-adatoms) in the region with Rashba coupling. 
We focus on the effects of the particular positions of such defects, i.e., their spatial 
distribution and their relative orientation with respect to the electric field. 

In principle, spin scattering is an important issue that may hinder the polarization of the current. Fortunately, the spin relaxation length in carbon systems  is rather large; for example, the experimental value reported for carbon nanotubes is around 50 $\mu$m \cite{Hueso2007}. 
Therefore, as long as the size of the device or system studied is much smaller than this length, spin coherence should be maintained and 
spin polarization could be achieved. This is certainly plausible in nanoscale devices. 

The article is organized as follows: In Sec. II we describe the systems, model and methods employed
to obtain the spin transport properties under a RSO interaction. 
Section III deals with the effects of isolated adatoms and  topological defects 
on the transport properties, the enhancement of spin-polarized currents 
due to multiple defects 
and the polarized-current dependence on their relative orientation;
finally, the interplay of RSO interaction and the effects of random 
H-adatoms are explored. 
Finally, in Sec. IV we summarize our main results.

\section{System and model}
\subsection{Description of the system}

\begin{center}
\begin{figure}[h]
\includegraphics[width=0.95\columnwidth]{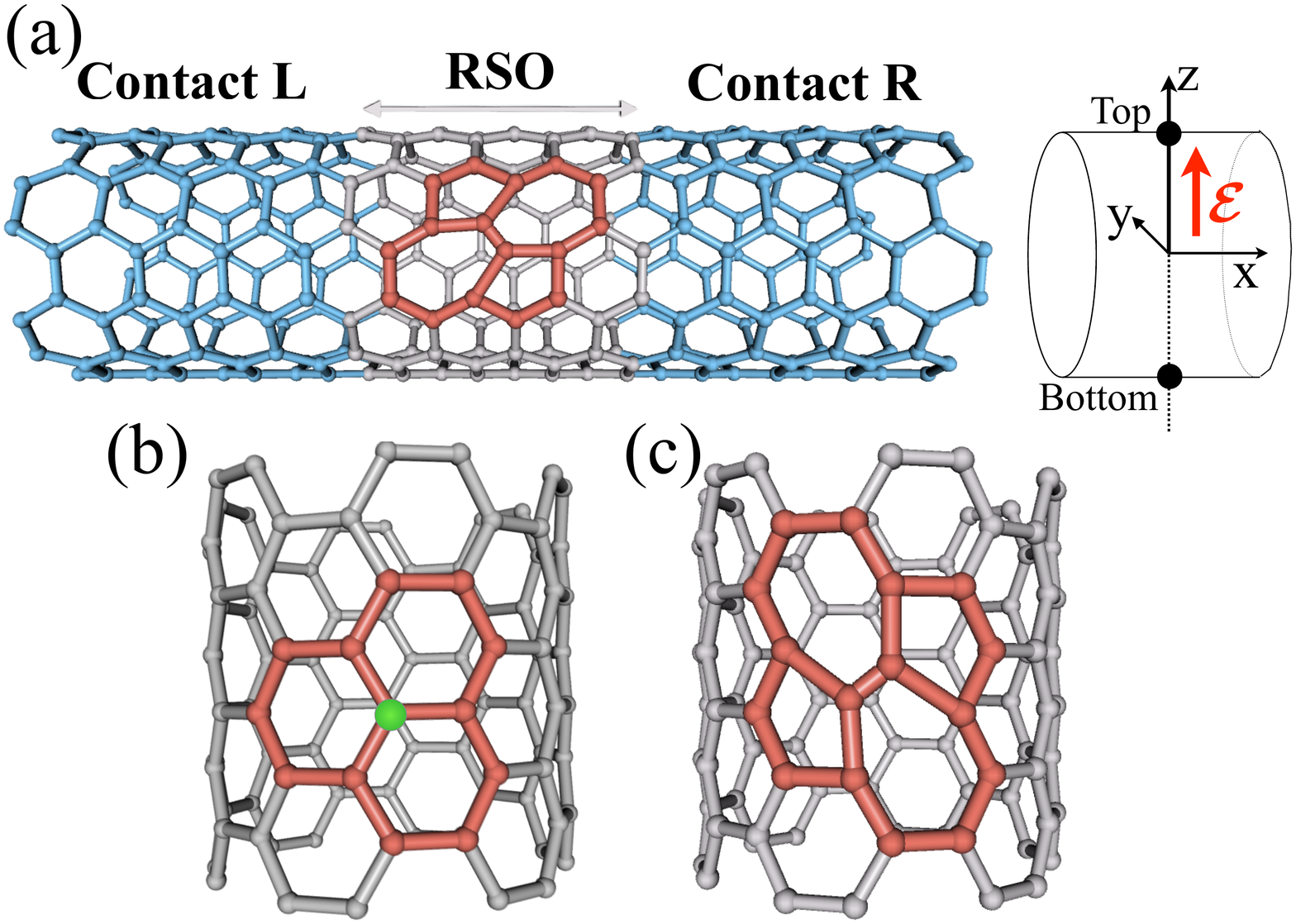}
\caption{\small (Color online)  (a) Schematic view of the device geometry. Left (L) and right (R) contacts are pristine CNTs without RSO interaction. 
The central part of the device is a conducting CNT of length $L$, 
with RSO interaction induced by the presence of an uniform electric field 
$\boldsymbol{\mathcal{E}}$ in the $z$ direction (red arrow).  
The directions $x$, ${y}$, and ${z}$ are shown in the right part of the 
figure; the electric field points into the $+z$ direction.  The top and bottom locations of the CNT are defined with respect to the 
electric field, as indicated. 
Examples of the defects considered (highlighted in red): 
(b) a H-adatom, shown in green on top of a carbon atom, and 
(c) a Stone-Wales topological defect.}
\label{fig1}
\end{figure}
\end{center}

To analyze the effects of localized defects on the spin-polarized currents 
generated by the Rashba spin-orbit interaction (SOI), we propose a device made 
of a CNT that has a region of length $L$ placed between two metallic 
plates with an uniform electric field 
$\boldsymbol{\mathcal{E}}$ 
between them [see Fig. \ref{fig1}(a)]. 
As we are interested in transport properties, we concentrate on metallic nanotubes, 
and restrict the study to armchair $(n,n)$ tubes. 
The central part under the applied electric field has an induced Rashba spin-orbit  (RSO) interaction; 
we name it  the \emph{Rashba region}.
Notice that the system may be described as an infinite carbon nanotube with three parts: a central finite Rashba region connected to two semi-infinite nanotube leads of the same chirality and radius \cite{HSantos2016}. 
The length $L$ of the Rashba region can be given in terms of the length $T$ of the translational vector by $L = N T$ (with $N \in \mathbb{N})$  \cite{Saito1992}. 
Recall that $T=\sqrt 3 a_c$ for  armchair tubes, where $a_{c}$ is the carbon bond length in graphene.

We introduce defects in this Rashba region. 
Specifically, as depicted in Fig. \ref{fig1}, 
we have chosen H-adatoms, shown in Fig. \ref{fig1}(b),
and 5-7-7-5 Stone-Wales (SW) defects 
(i. e., two carbon pentagon-heptagon pairs placed antiparallel to each other), rendered in Fig. \ref{fig1}(c). 
 The carbon atoms in which the defects produce important changes in the density of states 
 are highlighted in red, being 12 atoms for the H-adatom and 16 for the SW defect.

The presence of H-adatoms does not mix the two CNT sublattices, 
although depending on their number and position,  
they may break the sublattice symmetry, 
allowing the appearance 
of localized magnetic moments on the 
atoms 
close to them \cite{Palacios2008}. On the contrary, the introduction of topological defects in the hexagonal lattice of graphene or CNTs 
with an  
odd number of carbon atom rings  
mixes the two sublattices. 
Additionally, the combination of the two most probable topological defects in a carbon nanotube, 
a single  5-7 (pentagon-heptagon) pair, changes 
its chirality \cite{Chico1996a,Charlier1996}.
However, two adjacent 5-7 pairs in opposite direction, 
the 5-7-7-5 SW defect, preserves the chirality of the 
CNT, but there is still a local sublattice mixing. 
With the aim to simplify the analysis, we focus only on defects that 
preserve the chirality, and so we  study the effects of the RSO 
interaction on CNTs with H-adatoms and SW defects.

\subsection{Model and method}

Our calculation employs 
a single-orbital tight-binding model in the nearest-neighbor approximation 
with a RSO coupling limited to the central part of the carbon 
system \cite{Qiao2010,Lenz2013, Chico2015,HSantos2016}. 
The Hamiltonian can be written as 
$H=H_0 + H_R$, 
with the kinetic energy term being 
$H_0 = -\gamma_0 \sum c_{i\alpha}^\dagger  c_{j\alpha}$,
where $c_{i\alpha}^\dagger$ ($c_{j\alpha}$) 
is the creation (destruction) operator for electrons with 
spin projection $\alpha$ in site $i$ ($j$), and $\gamma_0$ 
is the nearest-neighbor hopping energy, which is constant 
throughout the system. This amounts to neglecting
the lattice relaxation produced by the defects, but this is a quantitative change 
with respect to lattice imbalance and mixing or orbital removal.

The RSO contribution to the Hamiltonian is 
\begin{equation} 
H_R = 
\frac{i \lambda_R}{a_{c}} 
\sum_{\substack {<i,j>\\\alpha,\beta}} c_{i\alpha}^\dagger  
\bigg[ 
(\boldsymbol{\sigma} \times \bold{d}_{ij}) \cdot \bold {e}_z 
\bigg]_{\alpha \beta} 
\  c_{j\beta} \,\,   ,
\label{HR} 
\end{equation}
where $\boldsymbol\sigma$ are the Pauli spin matrices, $\bold{d}_{ij}$ is the position vector between sites ${i}$ and ${j}$, and $\alpha, \beta$ are the spin projection indices. 
The unit vector $\bold{e}_z$ is along the electric field 
$\boldsymbol{\mathcal{E}}$ direction.
The Rashba spin-orbit strength  $\lambda_R$ is given by the 
electric field intensity, and its sign defined by the field orientation. 
A value of $\lambda_R = 0.1 \gamma_0$ is used \cite{Chico2015,HSantos2016}.
Although this value is very large with respect to the intrinsic spin-orbit interaction of a pristine 
carbon nanotube, the enhancements reported for functionalized and 
decorated graphene systems \cite{Marchenko2012,Balakrishnan2014} suggest that similar values may be achieved in carbon nanotubes with defects.

As previously discussed \cite{HSantos2016}, 
the direction for which the 
spin-polarized current is largest is that perpendicular to both the current and the external electric field; in our case it corresponds to the $y$ direction. 
This fact can be best understood by resorting to the expression of the Rashba term in 
the continuum approximation \cite{WinklerBook}, i. e, 
$H_R \propto (\boldsymbol{\sigma}  \times  \mathbf{k} ) \cdot \boldsymbol{\mathcal{E}} $, 
with $\mathbf{k}$ being the wave vector which is in the direction of the current.

The conductance along the nanotube axis ($x$ direction) is calculated following the Landauer approach in the zero bias approximation \cite{Chico1996a,HSantos2016}. 
Assuming unpolarized charge currents flowing from left to right, 
the spin-dependent conductance 
${G}^{LR}_{\sigma \sigma'}(E)$
is proportional to the probability that an electron with 
spin $\sigma$ and energy $E$ in the left contact 
reaches right contact with spin  $\sigma'$.
This probability can be calculated using the Green function 
formalism \cite{Xu2007,Diniz2012}. 
The spin polarization of the conductance in the $y$-direction is 
defined as 
\begin{equation} 
P_{y}(E) = 
G^{LR}_{\uparrow \uparrow} - G^{LR}_{\uparrow  \downarrow} + 
G^{LR}_{\downarrow \uparrow} - G^{LR}_{\downarrow \downarrow}\, .
\end{equation}
 As already mentioned, we restrict ourselves to the $y$ direction for the spin projections.  In the small-bias limit, this magnitude is proportional to the $y$-component of the spin current. 
Although other authors employ a normalized adimensional spin polarization \cite{Zhai2005,Diniz2012}, we have chosen this definition which is directly 
related to the spin-dependent conductances ${G}^{LR}_{\sigma \sigma'}(E)$.

The calculations are performed at zero temperature. We have checked that the effect of the temperature reduces $P_y$ and
gives rise to a smoother dependence with the energy, as expected, but 
the differences between the spin-polarized currents are still significant at room temperature. As an indicative value, we have found a maximum decrease of the polarization peaks of 60 \%. Such reduction does not preclude the possible measurement of the spin-polarized current and therefore, our results hold at room temperature. 

\section{ Results }

We consider an infinite metallic $(6,6)$ armchair CNT with a central 
finite Rashba region of varying length $L$,
which is given in terms of the length $T$ of the translational 
vector, $NT$.  
We consider Rashba regions with a varying density of defects, 
and we also investigate the optimal condition for the relative 
orientation between the defects and the direction of the electric field. 
 The Rashba interaction is switched on abruptly; we have 
verified that 
differences caused by considering a smooth profile for the electric field are irrelevant in the 
energy range of interest.  Likewise, we have checked that the position of the defects within the Rashba region,
i.e., with respect to its boundaries, is also unimportant in the energy range for which the bands are linear, around $E_F$.

We assume that an unpolarized current flows from the left semiinfinite 
nanotube through the Rashba region towards the right lead. 
In Ref. \cite{HSantos2016} we have shown that, for a pristine CNT, 
spin scattering in the Rashba region generates spin-polarized currents 
for certain spin directions and system symmetries. 
Here we analyze the effect of localized defects on the 
spin-polarized currents traversing the Rashba region.

As mentioned, we have focused on the analysis of the spin-projected 
conductance along the direction ($y$) perpendicular to both the current 
($x$) and the external electric field ($z$) directions, 
since it exhibits the largest spin-polarized current. 
In particular, 
we get zero spin polarizations $P_z$ and $P_x$  for all the armchair 
nanotubes $(n,n)$ with an even $n$ index for symmetry reasons 
\cite{HSantos2016}.
In contrast, the spin polarization of the conductance is maximized in the $y$ direction, 
for which there are no symmetry restrictions for its occurrence.

\subsection{ Single defects }

\begin{center}
\begin{figure}[h]
\includegraphics[width=1.\columnwidth]{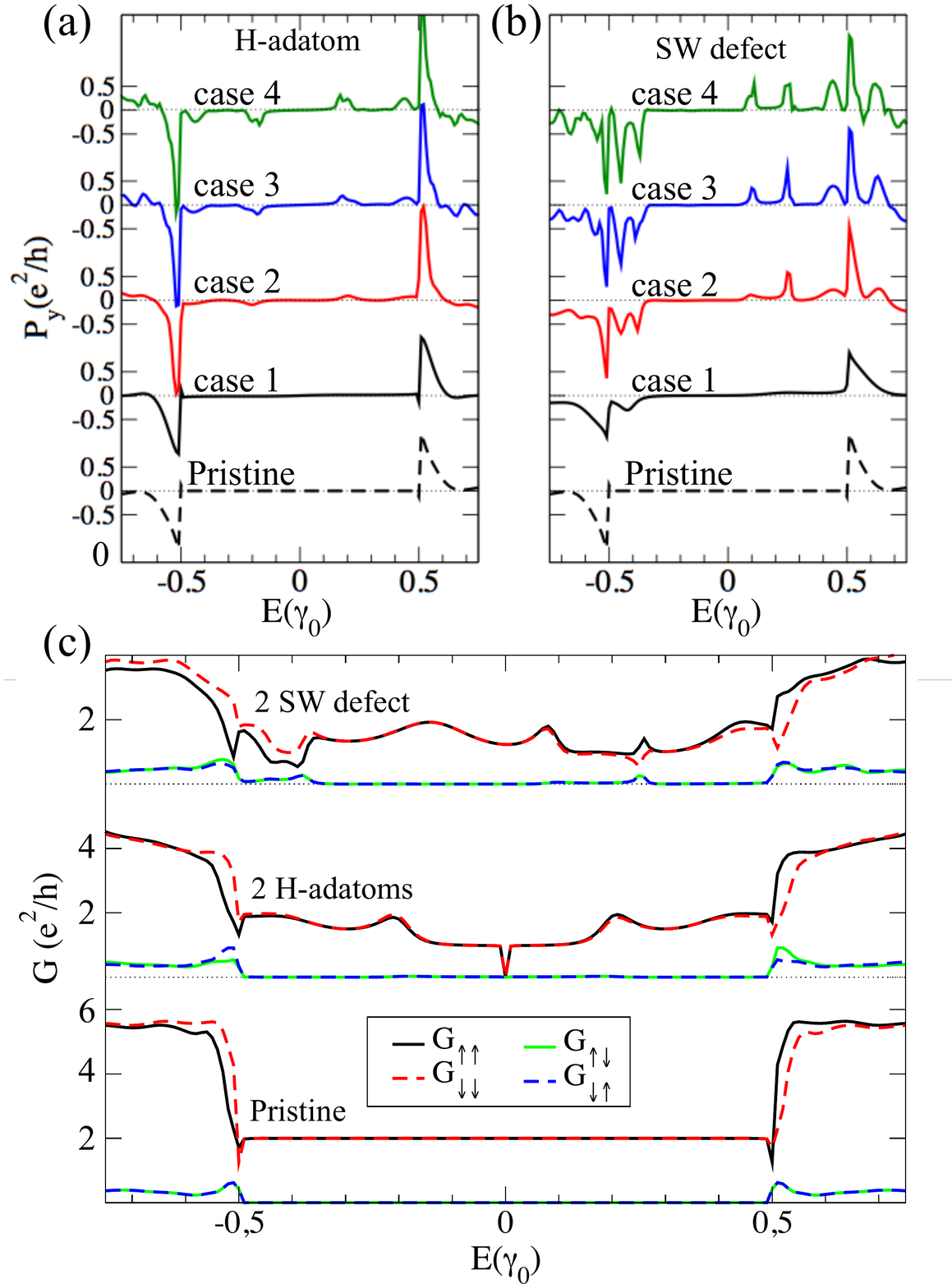}
\caption{(Color online) 
Armchair $(6,6)$ CNTs with a Rashba region with different lengths 
and number of defects. Panels (a) and (b): spin polarization of the conductance $P_y$. 
Each case $M$ corresponds to Rashba regions composed 
of $M$ finite CNT portions, each with a length $L_0 = 9 T$ 
and with a H-adatom or a SW defect. Panel (c): Spin-dependent conductances for a Rashba region 
$L = 2 L_0 = 18 T \approx 4.43$ nm. Low, middle, and upper graphs present curves for a pristine tube, 
two H-adatoms, and two SW defects equally spaced in the Rashba region, respectively. 
}
\label{fig2}
\end{figure}
\end{center}

The spin polarization of the conductances are evaluated for different sizes of the Rashba region, 
considering first the case of a single defect on the top position. 
Recall that we took the electric field direction to be $+z$, and 
the $x$ direction along the nanotube axis. 
We orient the nanotube so that the H-adatom
is situated on the $+z$ axis and the flux of the electric field 
through the area enclosed by the red highlighted bonds 
in Fig. \ref{fig1}(b) is maximum. 
For the SW defect, we proceed likewise, 
taking the $+z$ axis passing through the middle point 
of the bond between the two adjacent heptagons in Fig. \ref{fig1}(c). 
Note that both, the top and bottom positions, give the maximal 
flux of the electric field $\boldsymbol{\mathcal{E}}$ (in absolute value) through these 
defects.

Figures \ref{fig2}(a) and \ref{fig2}(b) present  
$P_y$   
for the pristine case and for the 
single defect case (case 1), 
using a Rashba region length $L_0 = 9 T \approx 2.21$ nm, 
both for the H-adatom and the SW defect. Notice that for the pristine nanotube, 
the spin polarization of the conductance is zero for energies close to the Fermi level, even though there are two 
channels available for the conductance related to the two different 
valleys \cite{Zhai2005}.
This is due to the lack of scattering processes in the Rashba region 
between states in 
different K valleys when the sole perturbation is the 
electric field \cite{Paul1999,HSantos2016}. 
But for energies close to $E = \pm 0.52 \gamma_0$, 
$P_y$ presents two peaks, allowed by the opening of other conductance channels 
in the nanotube contacts \cite{HSantos2016}. However, when a single defect (H-adatom or SW defect) 
is added to the Rashba region, intervalley scattering is allowed.
But the effect, as shown in Figs. \ref{fig2} (a) and (b), is too small to 
be noticeable for energies close to the Fermi level. 
These results for single defects are almost indistinguishable for the 
case of the H-adatom, whereas for the SW defect we observe 
small differences around peaks at  $E = \pm 0.52 \gamma_0$ 
[see case 1 in Figs. \ref{fig2}(a) and \ref{fig2}(b)]. 
Thus, when only one scattering center is considered in the Rashba region, 
the RSO effect close to the Fermi level is not easily distinguishable 
from the pristine case.

\subsection{Multiple defects}

In contrast to the single defect case, when we consider several $M$ 
equidistant defects at the top position of the Rashba region, the 
spin polarization of the conductance becomes significant close to the Fermi level. 
Previous works have analyzed the role of multiple defects in CNTs and graphene without taking RSO into account.
Due to the symmetry of the honeycomb lattice, there are certain distances between scattering centers that produce an unusual scattering behavior at zero energy \cite{GarciaLastra2008,Khalfoun2014,HSantos2014}: This happens if the defects in graphene or the unrolled nanotube sheet  are placed at a distance $\bold{R}=n \bold{a}_1 + m \bold{a}_2$ such that $m-n=3q$, with $q \in \mathbb{Z}$, where  $\bold{a}_1, \bold{a}_2$ are the graphene lattice vectors forming an angle of $60^{\rm o}$.  Such distances are associated with wave vectors that couple the two Dirac points K and K' at zero energy. But due to the linearity of the bands, intervalley coupling can be expected at other energies, provided that the bands are still linear. 
We choose different Rashba regions composed 
of $M$ $(6,6)$ finite CNT portions, 
each of them with a length $L_0$ and 
with a H-adatom or a SW defect  in it.
The defect density is kept constant, 
 i.e., one defect per length $L_0$. The top panels of Fig. \ref{fig2} [(a) and (b)] show the spin polarization 
of the conductance 
$P_y$ for the cases $M=1, 2, 3, 4$. 
The heights of the maxima with energies within the first conductance 
plateau range of the pristine CNT
increase with the number of defects. As for $L_0=9T$ the defects are located at a distance verifying the condition of the coupling of the 
two Dirac cones, the enhancement of the spin-conserved conductances is very strong. 

Such resonant behavior can be understood with a simple model for coherent scattering \cite{DattaBook}. Let ${\cal T}_1$ be the transmission probability through a single impurity. The transmission through two identical scatterers is given by 
${\cal T}_2= {\cal T}_1^2 /(1-2{\cal R}_1 \cos \delta + {\cal R}_1^2)$, 
where  ${\cal R}_1=1-{\cal T}_1$ is the reflection probability by the single defect and $\delta$ is the phase shift acquired in one round trip between the scatterers. At the resonance ($\cos \delta =1 $)  the total transmission ${\cal T}_2$ is always 1, independently of the value 
of  ${\cal T}_1$. In general, close to the resonance the transmission through the two defects ${\cal T}_2$ is larger than ${\cal T}_1$, so the presence of a second impurity at a distance fulfilling a resonant condition always yields an enhancement of the conductance.

The bottom panel, Fig. \ref{fig2} (c), represents the spin-dependent 
conductances as a function of the energy 
for the original pristine tube and the case $M=2$. 
The presence of the two defects, either H-adatoms or SW defects, is evident for the spin-dependent conductances $G_{\uparrow \uparrow}$ and $G_{\downarrow \downarrow}$ near $E_F$, being negligible for the spin-flip conductances $G_{\uparrow \downarrow}$ and $G_{\downarrow \uparrow}$.

For the nanotube with two H-adatoms at a distance $9T$, $P_y$ exhibits small peaks at energies 
$E =\pm 0.21\gamma_0$. 
The peaks coincide with the enhancement of the spin-dependent 
conductances $G_{\sigma \sigma}$ from $e^2/h$ to $2 e^2/h$ 
due to the electronic states localized around the 
adatoms \cite{HSantos2016}. 
This effect is more remarkable as the number of equispaced adatoms
in the Rashba region increases, as shown in Fig. \ref{fig2} (a) 
(cases 3 and 4). We have explained this increase as a consequence of resonant scattering. Small differences in the wavevectors for spin up and down transmission yield a non-zero $P_y$. 

More relevant changes in the spin-polarized currents appear for  the SW 
defects, as also shown in Fig. \ref{fig2} (b).
For case 2,  two defects separated by $9T$,  $P_y$ exhibits peaks at energies $E = 0.09 \gamma_0$ and 
$0.25 \gamma_0$;  
these peaks are higher for larger number of defects 
(cases 3 and 4). 
On the other hand, for energies around $E = \pm 0.5 \gamma_0$ 
(i.e., close to the transition from the first to the second conductance 
plateau of the pristine CNT leads) the $P_y$  for the SW defects exhibits many more 
features than for the H-adatoms.
Due to 
sublattice mixing, the electron-hole 
symmetry is broken for the tubes with SW defects, yielding a strong asymmetry in the results. 
On the contrary, 
$P_y$  for the nanotube with H-adatoms 
shows 
electron-hole symmetry,  $P_y(E)=-P_y(-E)$.

\subsubsection*{Band structure and spin analysis}

To understand the effect of H-adatoms and the SW defects in the 
spin-polarized currents induced by the Rashba SOI, we resort to the analysis of the spin-resolved band structure and the spin polarization of the conductance $P_y$ of an infinite periodic system. 
We choose a CNT unit cell of length $L_0$  with a single H-adatom or one SW defect in it, with  
 the RSO interaction present in the whole system.

\begin{center}
\begin{figure*}[!ht]
\includegraphics[width=0.9\textwidth]{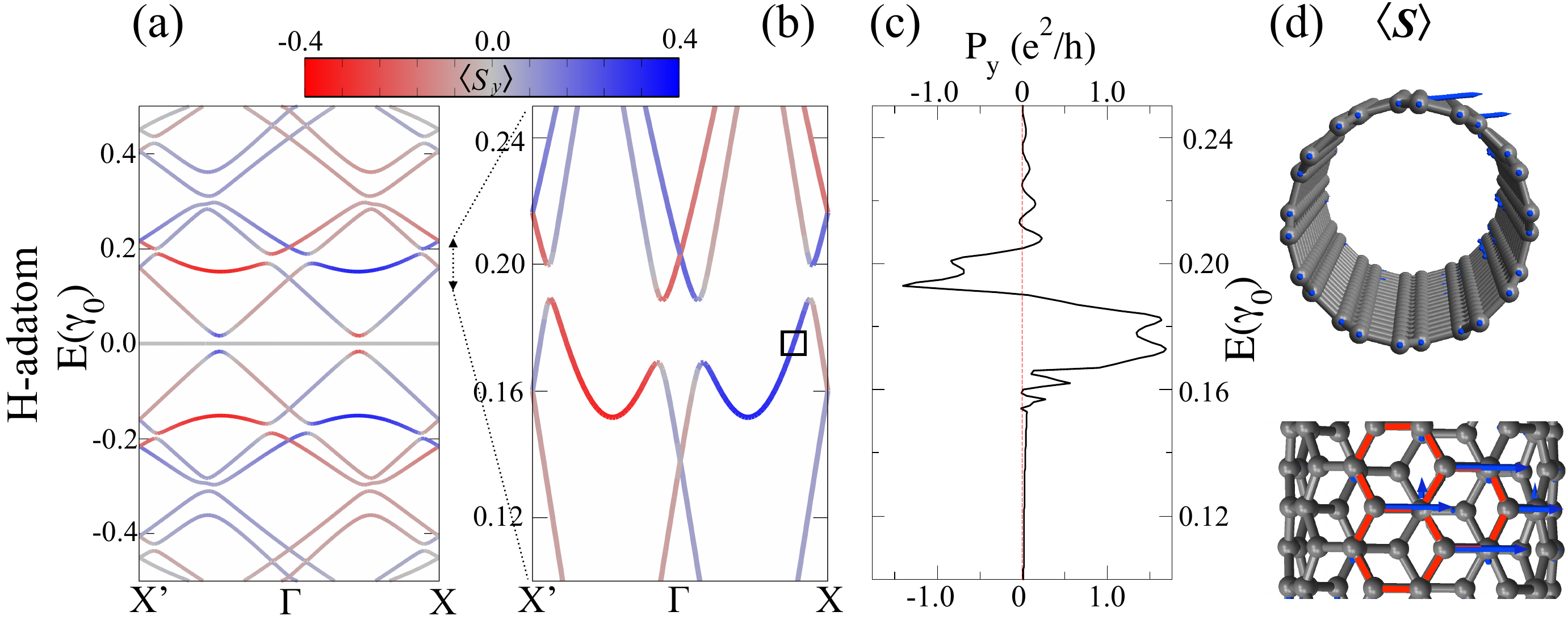}
\includegraphics[width=0.9\textwidth]{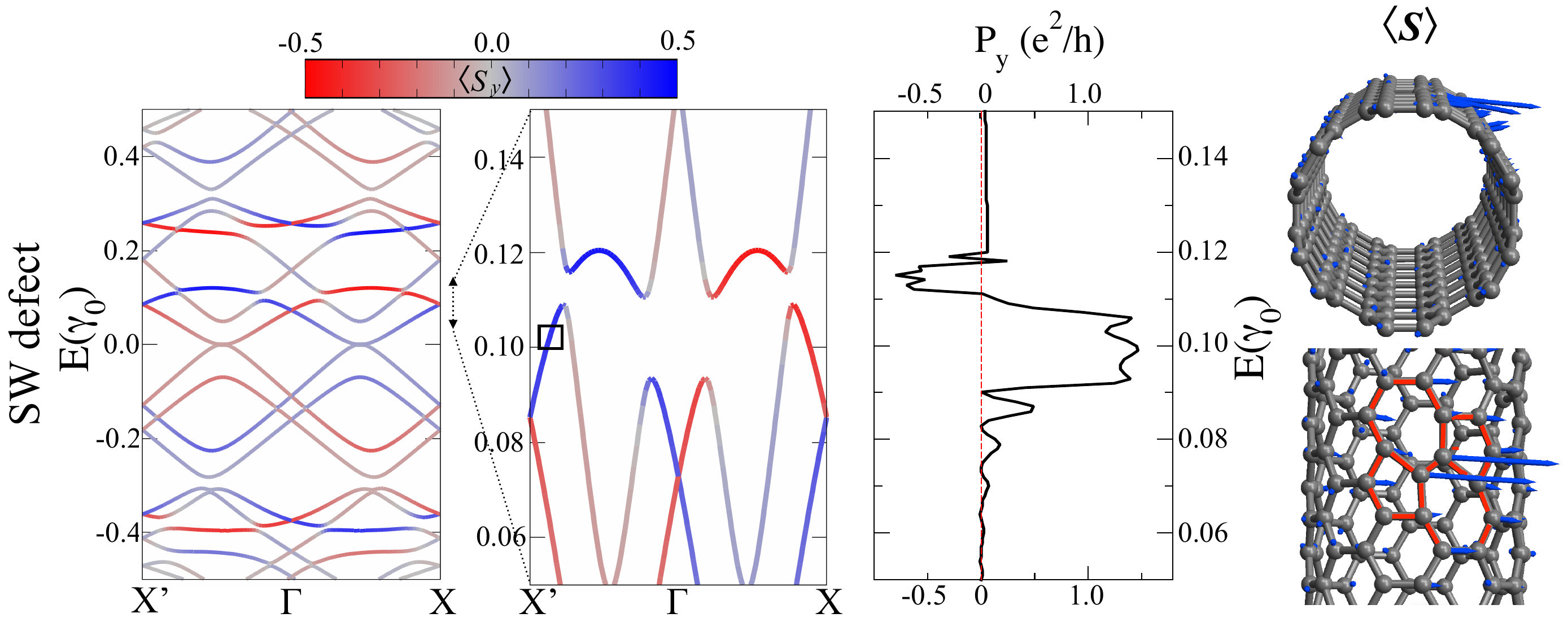}
\caption{(Color online) 
Results for the $(6,6)$ CNTs with H-adatoms or with SW defects 
under the RSO interaction. Panels (a): 
Spin-resolved band structures of infinite $(6,6)$ CNTs composed of finite 
nanotube stacks, each of a length $L_0 = 9 T$ and with one defect.
Zooms of the band structures in the relevant energy range are shown 
in the panels (b). The expectation values of the spin component $\langle S_y \rangle $ at the adatoms and SW defects, as defined in Fig. \ref{fig1}, are presented in a color scale (blue for up spin and red for down spin). Panels (c): 
The corresponding $P_y$ are shown for 
$L=20 \, L_0$, where $L_0=9T \approx 2.21$ nm. Panels (d):  Lateral schematic view of the nanotube with a H-adatom and with a SW defect. Blue arrows denote $\langle \mathbf{S} \rangle $ values at atoms in the defect for the $k$ values highlighted with black squares in panels (b).}
\label{fig3}
\end{figure*}
\end{center}

Figure \ref{fig3}(a) depicts the band structures for the 
$(6,6)$ CNT with Rashba SOI, with either H-adatoms (upper panel) 
or SW defects (bottom panel). The expectation values for the $y$ component of the spin, 
$\langle S_y \rangle $, projected in all the atoms around the defect or belonging to it  
 (highlighted in red in Fig. \ref{fig1}), are shown with a color scale in the bands. 
The expectation value of the spin in the $y$ direction at a specific atom $n$ is given by  
$\langle S_y \rangle_n = \frac{\hbar}{2} \langle \sigma_y \rangle_n$, with 
$\langle \sigma_y \rangle_n = \langle \phi_n  \vert  \sigma_y \vert \phi_n \rangle$, 
where $\phi_n$ is the tight-binding atomic eigenfunction in 
the single-orbital model. We have also plotted in Fig. \ref{fig3} a zoom of the band structure 
in the relevant energy range [Fig. \ref{fig3}(b)] 
as well as the corresponding  spin polarizations of the conductances $P_y$ [in Fig. \ref{fig3}(c)].

The bands of the periodic (6,6) CNT of length $9T$ with one H-adatom present large absolute values of $\langle S_y \rangle$ in the energy ranges from $\pm 0.16 \gamma_0$ to $ \pm 0.22 \gamma_0$, 
coinciding with the bumps of $P_y$ in Fig. \ref{fig2}(a) 
(with $L$ up to $4 L_0$, where $L_0=9T \approx 2.21$ nm).
These peaks in $P_y$ are much clearer in Fig. \ref{fig3}(c), obtained for a system with $L=20 L_0$. 
Recall that for current 
flowing from left to right, the states that contribute to the current 
scattering process are those with positive velocity 
\cite{DattaBook, SCB09}. 
The zoom in the top panel of Fig. 3(b) shows a band with a strong positive $\langle S_y \rangle $
  from   $ 0.16\,\gamma_0$ to $0.19\,\gamma_0$ 
in the wavevector interval  $k= 0.25\,\overline{\Gamma \mathrm{X}}$ 
to $k = 0.40\,\overline{\Gamma \mathrm{X}}$. 
However, there is no band with negative  $\langle S_y \rangle $ in the same energy range, due to the opening of the gap around $\Gamma$ 
by the strong hybridization. 
At the same energy range $P_y$ shows a positive peak 
(Fig. \ref{fig3}(c), top). 
Besides, 
$P_y$ shows a negative peak from 
$0.195\,\gamma_0$ to $0.21\,\gamma_0$, because the band with positive slope 
in that energy interval has negative $\langle S_y \rangle $. 

The same analysis can be performed for CNTs with topological defects. 
The spin-resolved band structure for a periodic (6,6) nanotube with a SW defect is presented in the bottom panel of 
Fig. \ref{fig3} (a). Large absolute values 
of $\langle S_y \rangle $ occur 
from $ 0.08\,\gamma_0$ to $0.12\,\gamma_0$ 
and from $ 0.24\,\gamma_0$ to $0.26\,\gamma_0$. As in the H-adatom case, the spin-polarized current appears when there is an inequality between the
number of channels with opposite values of $\langle S_y \rangle $ due to the gap opening. 
 These energy ranges coincide with the $P_y$ enhancements shown in Fig. \ref{fig2}(b). 

From these results we can assume that the states producing a spin polarization of the conductance are associated with the defects. To verify this assumption, we plot in 
Figs. \ref{fig3} (d)  the expectation values of the spin $\langle \mathbf{S} \rangle$ 
corresponding to the states which give rise to the polarized current at the atoms around (H-adatom) or composing (SW) the defect. 
Their wavevectors $k$ are marked with black squares in the zooms of the band structures presented in  
Figs. \ref{fig3} (b).
This expectation value is projected on each atom $n$, 
$\langle \mathbf{S} \rangle_n = 
\langle S_x \rangle_n \ \mathbf{e}_x + 
\langle S_y \rangle_n  \ \mathbf{e}_y + 
\langle S_z \rangle_n  \ \mathbf{e}_z $, 
where 
$\langle S_i \rangle_n = \frac{\hbar}{2} \langle \sigma_i \rangle_n$, 
$i=x$, $y$, $z$, 
and 
$\langle \sigma_i \rangle_n = \langle \phi_n  \vert  \sigma_i \vert \phi_n \rangle$.
Their energies and wavevectors are $E=0.173\,\gamma_0$ and $k = 0.375 \ \overline{\Gamma \mathrm{X}}$ for the H-adatom and $E = 0.1065 \, \gamma_0$ and $k = 0.41 \ \overline{\Gamma \mathrm{X}'}$ for the SW defect. 
The large absolute values of $\langle \mathbf{S} \rangle$ in those bands are located at the defects; thus, they are related to the high spin polarization of the conductances presented 
in Fig. \ref{fig3} (c).

Taking into account the features of the bands shown for the periodic CNT with defects (Fig. \ref{fig3}), we conclude that the peaks of $P_y$ around $E_F$  are actually associated with the existence of equispaced defects in the Rashba region.
It is important to note that besides the spin polarization of the bands, gaps should open at the same energies for other wavevectors.  In this way, an imbalance between the channels with opposite values of the spin projection is produced, giving rise to spin-polarized currents.

It is noteworthy that if the distance between nearest defects does not fulfill the resonant condition ($N \neq 3q$), after four or more repetitions an increase of the spin polarization of the current is produced anyway. Obviously, for $M=4$, there are two impurities at a distance that verifies such condition, so the effect is always reinforced for periodically located defects. To illustrate this point, 
we present in Figs. \ref{fig4} and \ref{fig5} the cases of 20 equispaced H-adatoms and SW defects, respectively, on the top positions (A) with $L_0 = 4T$, showing quite strong peaks at similar energy ranges.

\subsection{Influence of the relative alignment of defects and the electric field}
 
\begin{center}
\begin{figure}[t]
\includegraphics[width=1.0\columnwidth]{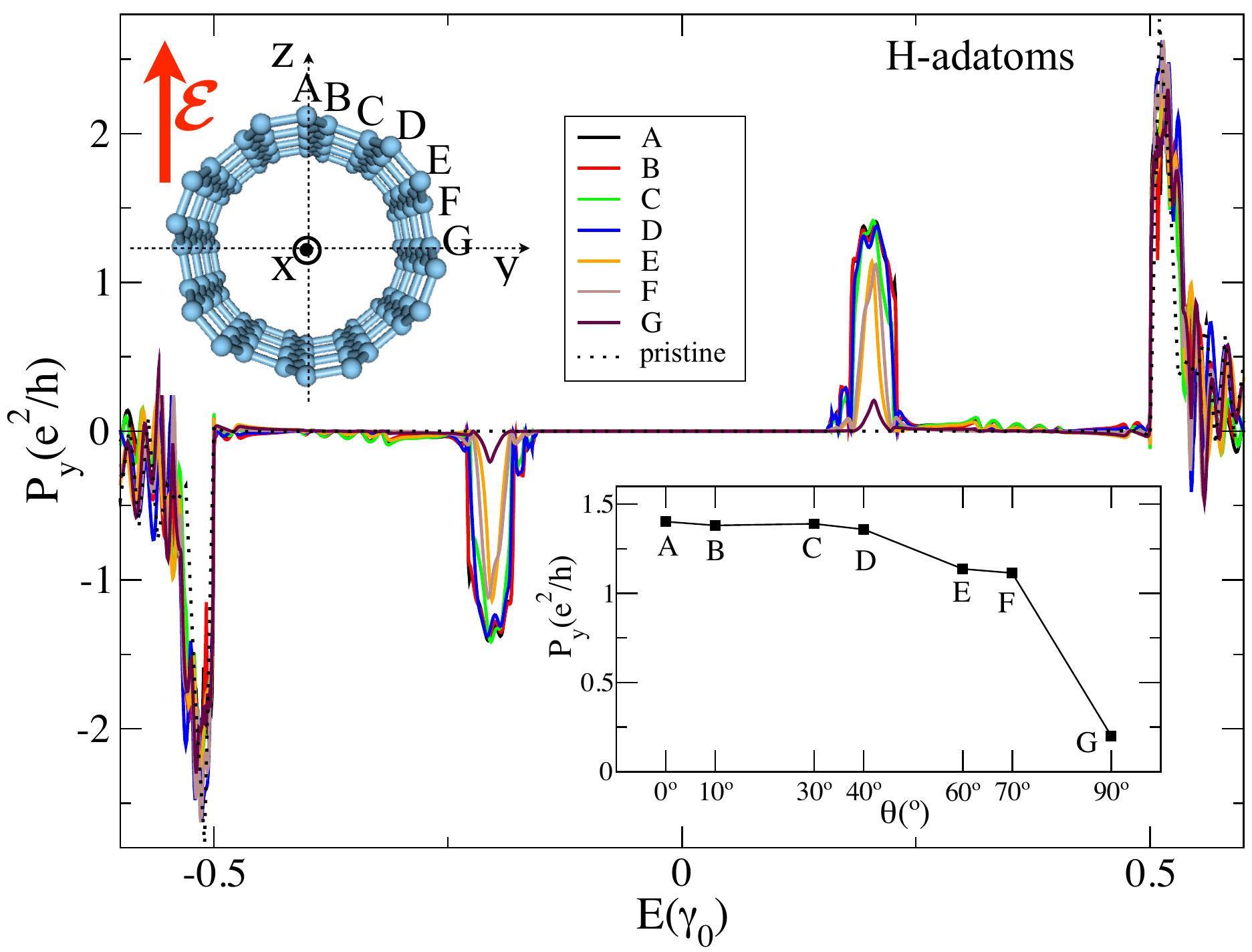}
\caption{(Color online) 
Spin polarization of the conductance, $P_y$, for a  $(6,6)$ CNT with a Rashba 
region composed of 20 H-adatoms equally spaced in a length given by 
$20 \, L_0$ ($L_0=4T\approx 0.98$ nm),
for different positions of the adatom relative to the electric field direction.  Position A corresponds to all the H-adatoms placed at the top of the tube, whereas G when the adatoms are situated at the side of the tube. Intermediate cases (B to F) are also investigated.  All H-adatom configurations are graphically detailed in the CNT scheme 
at the top inset.}
\label{fig4}
\end{figure}
\end{center}

\begin{center}
\begin{figure}[h]
\includegraphics[width=1.0\columnwidth]{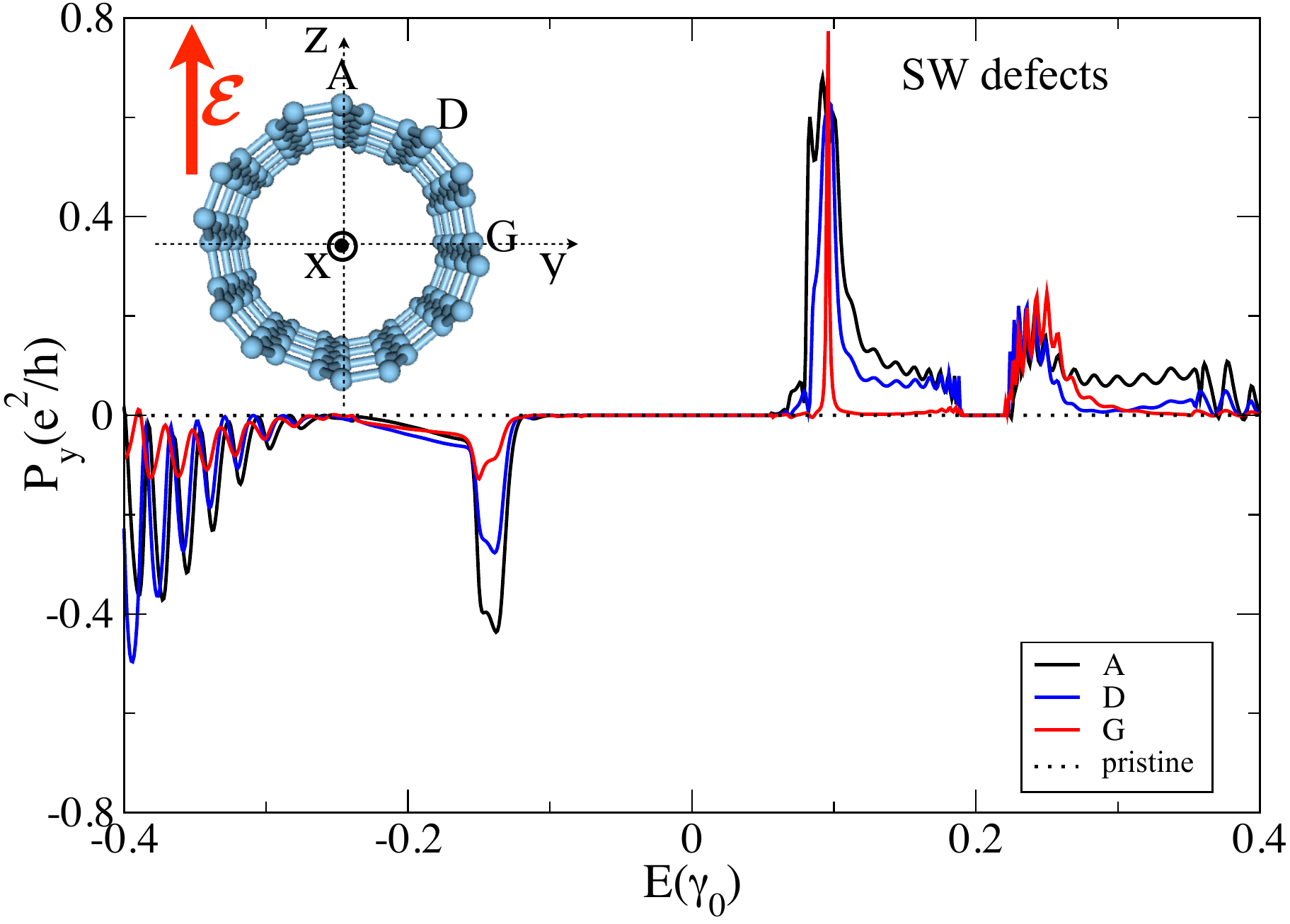}
\caption{(Color online) 
Spin polarization of the conductance, $P_y$, 
for a $(6,6)$ CNT with a Rashba region length equal to 
$20 \, L_0$ ($L_0=4T \approx 0.98$ nm), with 20 SW defects inside,  for different positions of the defects with respect to the electric field.  Position A corresponds to all the defects located at the top of the tube,  G when they are situated at the  nanotube side and D is an intermediate case.  
All the three cases are detailed in the CNT scheme at the top inset.}
\label{fig5}
\end{figure}
\end{center}

Here we analyze the dependence of the spin-polarized currents on the relative position of the defect in the original tube with respect to the direction of the electric field.
We consider a 
(6,6) CNT with a Rashba region composed of $20 \ L_0$  with their corresponding adatoms equally spaced 
with  $L_0 = 4 \, T$, representing a sample with higher density of defects 
within the Rashba area. With the electric field along the $+z$ axis,  the relative 
orientation of the defects is varied from the parallel direction with 
respect to the field 
(top position, labeled A in Figs. \ref{fig4} and \ref{fig5}) 
to the perpendicular direction (lateral side, labeled G), 
through intermediate positions. In Fig. \ref{fig4} we present the results of the spin polarization 
 of the conductance $P_y$ for the H-adatom, considering the 
positions in the tube shown in the inset of the figure. 
The peaks of $P_y$ close to the Fermi energy 
decrease in intensity as the 
H-adatom positions change from  A to G, with a sharp reduction of $P_y$ from F to G. This is highlighted in the inset  of Fig. \ref{fig4}, 
which shows the dependence of the maximum value of $P_y$ 
on the angle between the electric field and the normal to the defects. 
Such variation can be expected from the expression of the Rashba Hamiltonian: at position A, the vectors spanning the bonds in the defects are 
perpendicular to $\boldsymbol{\mathcal{E}}$, 
maximizing their contribution to the Rashba term, whereas for position G the electric field is contained in the plane 
formed by these bonds, yielding a much smaller contribution. As mentioned, the electric flux of $\boldsymbol{\mathcal{E}}$
through the defects is maximal when the defects are at the top 
or the bottom positions on the nanotube. 
Therefore, the effect of the RSO is maximized for a maximum flux of the electric field through the defects.

Similar results are found for the SW defects. Figure \ref{fig5} shows $P_y$ for three relative orientation of the defects 
in the tube with respect to the electric field direction, indicated in 
the inset of the figure (positions A, D and G). The same defect density and length of the Rashba region employed for the H-adatoms are used here for the SW defects. 
Going from A to G, two noticeable effects can be observed. First, we see a large decrease of the peak at $E=-0.14\,\gamma_0$. 
In fact,  the height of this peak  
when the defects are located at G (the lateral side of the tube) is only about $30\%$ of the value obtained for the top position A. 
Second, we observe a pronounced narrowing of the width of the peak 
that appears in the  energy range  between $E=0.1\,\gamma_0$ and $0.18\,\gamma_0$.  As expected, the relative position of the defects with respect to the electric field plays an important role in the spin polarization of the conductance.

\begin{center}
\begin{figure}[h]
\includegraphics[width=1.0\columnwidth]{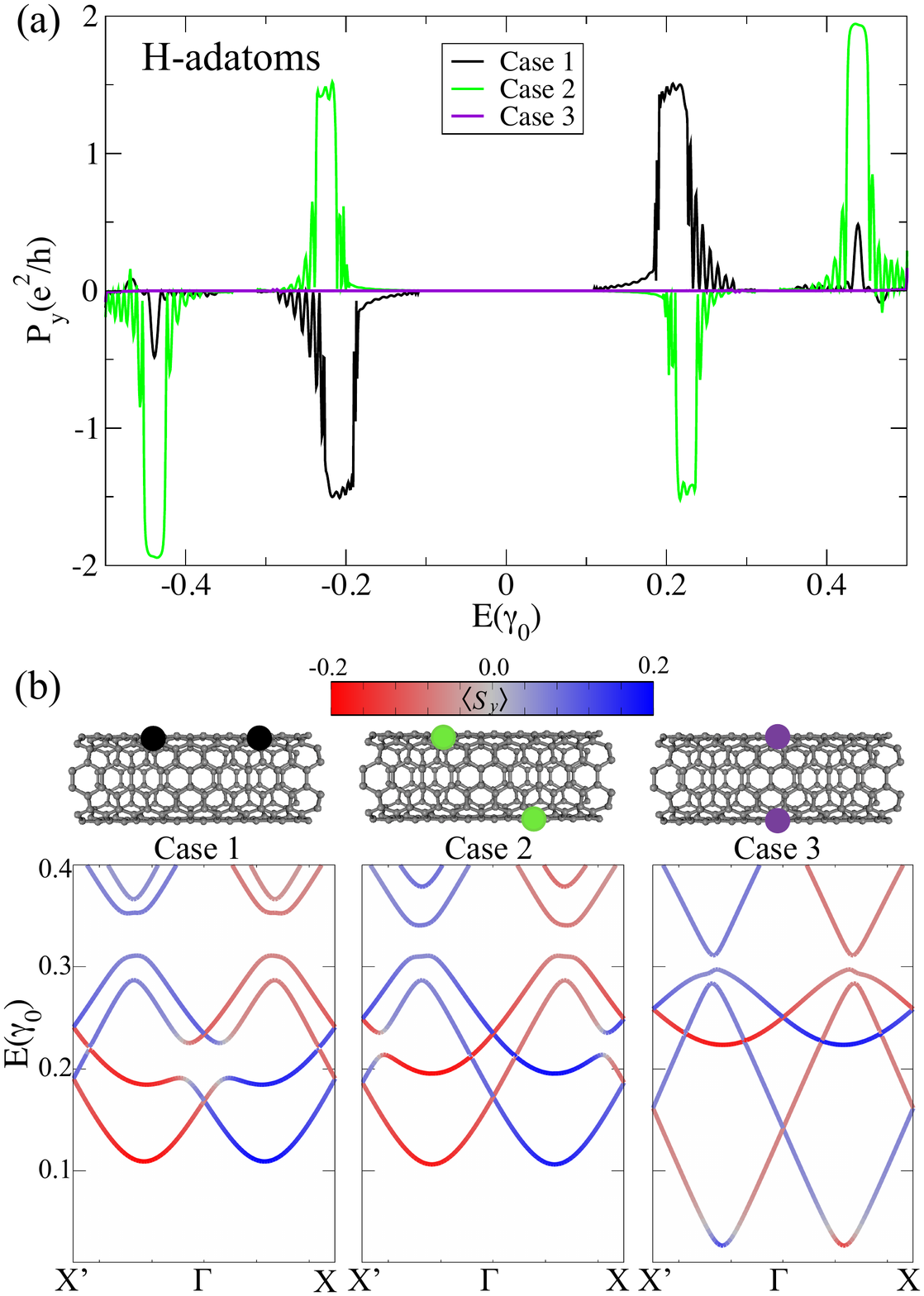}
\caption{(Color online) 
(a) Spin polarization of the conductance, $P_y$, for a $(6,6)$ CNT with a Rashba region of 
length $20 \, L_0$ ($L_0=9T\approx 2.21$ nm) with two H-adatoms in $L_0$. 
Three configurations with different relative positions between the adatoms are presented, 
as detailed in the scheme; the colored dots indicate the position of the H-adatoms.
(b) Spin-resolved electronic band structures for the three cases, with  $\langle S_y \rangle $ shown in a color scale.
  }
\label{fig6}
\end{figure}
\end{center}

To investigate the effects of the relative positions of the defects, we further consider the case of two H-adatoms in the unit 
length $L_0$, taking a Rashba region length equal to $20 \, L_0$, 
where $L_0$ is $L_0=9 \, T$. 
In case 1, two H-adatoms are both placed at the top of the tube.
In case 2, one H-adatom is at the top and the other at the bottom of the tube, 
but not in diametrically opposite sites of the tube. 
In case 3 
the H-adatoms are also at the top and the bottom of the tube, but exactly at opposite sites, in a symmetric configuration.  
The corresponding $P_y$ are shown in Fig. \ref{fig6} (a). 
Again, for all cases we obtain two maxima  of  
$P_y$ close to energies $\pm 0.2 \, \gamma_0$.
It is noteworthy that $P_y$ changes depending on the relative position 
of the H-adatoms in the unit slab $L_0$. 
For case 2 the peaks of $P_y$ are narrower than those of case 1,  and their signs are reversed. 
For the symmetric case 3, the polarization  completely disappears 
in the region corresponding to the first conductance plateau, identical to the results obtained for the pristine case. 

To gain some insight into the main different features related to the 
relative position of the adatoms, 
we also resort to the band structures of infinite (6,6) CNTs with 
the same defect distributions showing the spin expectation value  
$\left\langle S_y \right\rangle$
of the states in the bands, 
 depicted in Fig. \ref{fig6} (b). 
 They present dramatic differences. The bands of case 1 show large anticrossings at $\Gamma$ around the energies $\pm 0.2 \, \gamma_0$, thus leaving unbalanced conduction channels 
with positive $\langle S_y \rangle$ near X and X'. 
Case 2 shows the opposite behavior: the anticrossings take place near the X and X' points, so the remaining channels close to $\Gamma$ have a negative 
spin projection $\left\langle S_y \right\rangle$. 
As a result, in case 2 the peaks of $P_y$ have inverted signs compared to case 1. 
In the symmetric case 3, polarizations and slopes of the bands are 
completely compensated. Since there are not any anticrossings nor band openings 
at these energies, the spin polarization of the conductances is zero, despite the fact that the bands show net spin expectation values.

 These results can be of practical interest. The fact that adatoms situated 
 on the same side of the tube give rise to 
spin-polarized currents that present inverted polarization signals when the 
H-adatoms are situated in opposite sides (top and bottom sides of the tube), 
may be of great importance, for instance, for transport and storage of spin information.
Actually, DNA molecules wrapped around single-walled carbon nanotubes are shown to 
induce a helicoidal electric field due to the polar nature of the adsorbed DNA molecule 
\cite{Diniz2012} and spin-polarized currents are produced 
depending on the symmetry of the DNA molecule-nanotube hybrid system. 
Our results show that it could be possible to engineer particular defect 
configurations on the CNTs, opening new routes for 
their use in spintronic devices \cite{Alam2015}.

\begin{center}
\begin{figure}[h]
\includegraphics[width=1.0\columnwidth]{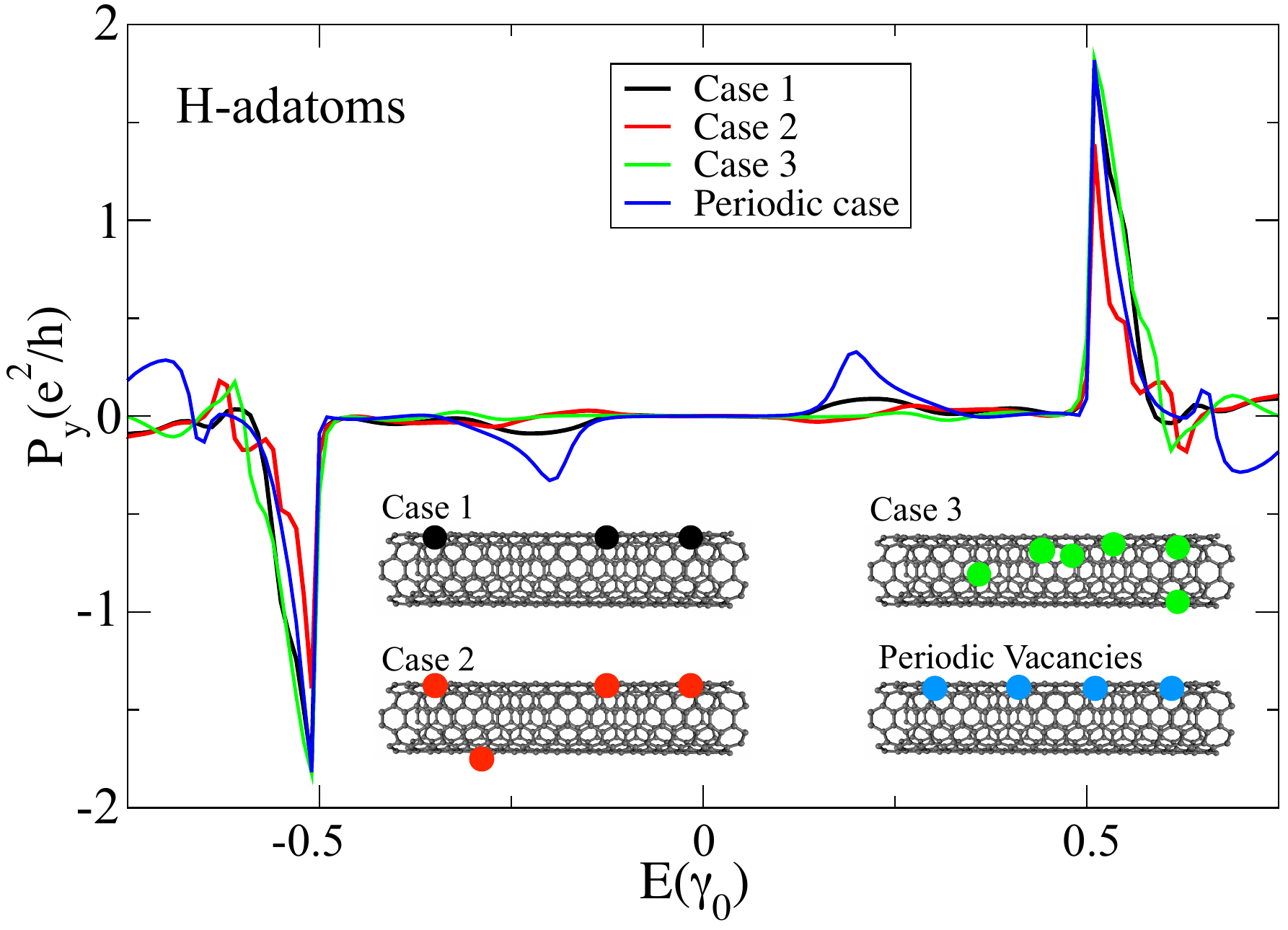}
\caption{(Color online) 
Spin polarization of the conductance $P_y$ for a $(6,6)$ CNT with a Rashba region composed 
of 16 unit cells ($L_0=16 \, T$) for the distributions of H-adatoms given in the 
insets.}
\label{fig7}
\end{figure}
\end{center}

\subsection{Irregularly spaced H-adatoms}

Finally, to extend our analysis beyond 
periodic configurations of defects, we have also considered several unequally spaced H-adatom distributions within the Rashba region (see the insets in Fig. \ref{fig7}). 
In cases 1 and 2 we have three and four H-adatoms in top 
and bottom positions with different spacings. 
Seven adatoms at random positions have been used in case 3.
Four equispaced H-adatoms have also been calculated for 
comparison. In all cases, the length of the Rashba region is taken as 
$L_0=16 \, T \approx 3.94$ nm.

Figure \ref{fig7} shows $P_y$ for 
these different adatom distributions. 
For cases 1 to 3 the height of those peaks closer to the Fermi 
energy is clearly smaller than for the equispaced defects. 
Therefore, it appears that the best scenario to obtain spin-polarized 
conductances in these CNTs is when the H-adatoms are in a periodic arrangement
within the Rashba region.

\section{Summary}

We have presented a detailed study of the effects of H-adatom and 
Stone-Wales defects on the  spin-polarized currents induced by a 
RSO interaction in metallic carbon nanotubes. 
Our main results show that 
(i) defects produce an increase of the spin-polarized currents at energies 
close to the  Fermi energy; 
(ii) spin-polarized currents appear 
when an imbalance between the number of spin-polarized conductance channels with opposite spin direction 
takes place at the same energy,  and furthermore 
(iii) such currents are greatly enhanced when the defects are  periodically introduced
in the system.  
Actually, a single defect scenario is insufficient 
to generate significant spin-dependent currents, 
but adding a few equispaced defects 
yields significant results. 
In summary,  introducing a regular array of defects in the Rashba region is optimal for 
the obtention of spin-polarized currents in carbon materials. This can be explained as a consequence of resonant tunneling through the multiple defects.

We have also found the importance of the orientation of the defects 
with respect to the direction of the electric field that causes the RSO 
interaction. 
Specifically, we have observed that a change in the relative position of 
neighbor defects (from top-top to top-bottom) produces a reversal of the 
 spin polarization $P_y$, which can be exploited as a spin switch. 
Our results can be useful for exploring new routes in the design 
of spintronic devices with CNTs.

\section{Acknowledgments}

A. L. acknowledges the financial support of FAPERJ through Grants No. 
E-26/102.272/2013 and No. E-26/202.953/2016, CNPq, and INCT em Nanomateriais de Carbono.
L. C. acknowledges support from Spanish MINECO through Grant No. FIS2015-64654-P, and helpful conversations with Jorge I. Cerd\'a. 
H. S. is grateful for financial support from the Brazilian CAPES.


\begin{thebibliography}{32}%
\makeatletter
\providecommand \@ifxundefined [1]{%
 \@ifx{#1\undefined}
}%
\providecommand \@ifnum [1]{%
 \ifnum #1\expandafter \@firstoftwo
 \else \expandafter \@secondoftwo
 \fi
}%
\providecommand \@ifx [1]{%
 \ifx #1\expandafter \@firstoftwo
 \else \expandafter \@secondoftwo
 \fi
}%
\providecommand \natexlab [1]{#1}%
\providecommand \enquote  [1]{``#1''}%
\providecommand \bibnamefont  [1]{#1}%
\providecommand \bibfnamefont [1]{#1}%
\providecommand \citenamefont [1]{#1}%
\providecommand \href@noop [0]{\@secondoftwo}%
\providecommand \href [0]{\begingroup \@sanitize@url \@href}%
\providecommand \@href[1]{\@@startlink{#1}\@@href}%
\providecommand \@@href[1]{\endgroup#1\@@endlink}%
\providecommand \@sanitize@url [0]{\catcode `\\12\catcode `\$12\catcode
  `\&12\catcode `\#12\catcode `\^12\catcode `\_12\catcode `\%12\relax}%
\providecommand \@@startlink[1]{}%
\providecommand \@@endlink[0]{}%
\providecommand \url  [0]{\begingroup\@sanitize@url \@url }%
\providecommand \@url [1]{\endgroup\@href {#1}{\urlprefix }}%
\providecommand \urlprefix  [0]{URL }%
\providecommand \Eprint [0]{\href }%
\providecommand \doibase [0]{http://dx.doi.org/}%
\providecommand \selectlanguage [0]{\@gobble}%
\providecommand \bibinfo  [0]{\@secondoftwo}%
\providecommand \bibfield  [0]{\@secondoftwo}%
\providecommand \translation [1]{[#1]}%
\providecommand \BibitemOpen [0]{}%
\providecommand \bibitemStop [0]{}%
\providecommand \bibitemNoStop [0]{.\EOS\space}%
\providecommand \EOS [0]{\spacefactor3000\relax}%
\providecommand \BibitemShut  [1]{\csname bibitem#1\endcsname}%
\let\auto@bib@innerbib\@empty
\bibitem [{\citenamefont {Padilha}\ \emph {et~al.}(2011)\citenamefont
  {Padilha}, \citenamefont {Amorim}, \citenamefont {Rocha}, \citenamefont
  {da~Silva},\ and\ \citenamefont {Fazzio}}]{Padilha2011}%
  \BibitemOpen
  \bibfield  {author} {\bibinfo {author} {\bibfnamefont {J.}~\bibnamefont
  {Padilha}}, \bibinfo {author} {\bibfnamefont {R.}~\bibnamefont {Amorim}},
  \bibinfo {author} {\bibfnamefont {A.}~\bibnamefont {Rocha}}, \bibinfo
  {author} {\bibfnamefont {A.}~\bibnamefont {da~Silva}}, \ and\ \bibinfo
  {author} {\bibfnamefont {A.}~\bibnamefont {Fazzio}},\ }\href {\doibase
  http://dx.doi.org/10.1016/j.ssc.2010.12.031} {\bibfield  {journal} {\bibinfo
  {journal} {Solid State Communications}\ }\textbf {\bibinfo {volume} {151}},\
  \bibinfo {pages} {482 } (\bibinfo {year} {2011})}\BibitemShut {NoStop}%
\bibitem [{\citenamefont {McCreary}\ \emph {et~al.}(2012)\citenamefont
  {McCreary}, \citenamefont {Swartz}, \citenamefont {Han}, \citenamefont
  {Fabian},\ and\ \citenamefont {Kawakami}}]{Kathleen2012}%
  \BibitemOpen
  \bibfield  {author} {\bibinfo {author} {\bibfnamefont {K.~M.}\ \bibnamefont
  {McCreary}}, \bibinfo {author} {\bibfnamefont {A.~G.}\ \bibnamefont
  {Swartz}}, \bibinfo {author} {\bibfnamefont {W.}~\bibnamefont {Han}},
  \bibinfo {author} {\bibfnamefont {J.}~\bibnamefont {Fabian}}, \ and\ \bibinfo
  {author} {\bibfnamefont {R.~K.}\ \bibnamefont {Kawakami}},\ }\href {\doibase
  10.1103/PhysRevLett.109.186604} {\bibfield  {journal} {\bibinfo  {journal}
  {Phys. Rev. Lett.}\ }\textbf {\bibinfo {volume} {109}},\ \bibinfo {pages}
  {186604} (\bibinfo {year} {2012})}\BibitemShut {NoStop}%
\bibitem [{\citenamefont {Ugeda}\ \emph {et~al.}(2010)\citenamefont {Ugeda},
  \citenamefont {Brihuega}, \citenamefont {Guinea},\ and\ \citenamefont
  {G\'omez-Rodr\'{\i}guez}}]{Ugeda2010}%
  \BibitemOpen
  \bibfield  {author} {\bibinfo {author} {\bibfnamefont {M.~M.}\ \bibnamefont
  {Ugeda}}, \bibinfo {author} {\bibfnamefont {I.}~\bibnamefont {Brihuega}},
  \bibinfo {author} {\bibfnamefont {F.}~\bibnamefont {Guinea}}, \ and\ \bibinfo
  {author} {\bibfnamefont {J.~M.}\ \bibnamefont {G\'omez-Rodr\'{\i}guez}},\
  }\href {\doibase 10.1103/PhysRevLett.104.096804} {\bibfield  {journal}
  {\bibinfo  {journal} {Phys. Rev. Lett.}\ }\textbf {\bibinfo {volume} {104}},\
  \bibinfo {pages} {096804} (\bibinfo {year} {2010})}\BibitemShut {NoStop}%
\bibitem [{\citenamefont {Gonz{\'a}lez-Herrero}\ \emph
  {et~al.}(2016)\citenamefont {Gonz{\'a}lez-Herrero}, \citenamefont
  {G{\'o}mez-Rodr{\'\i}guez}, \citenamefont {Mallet}, \citenamefont {Moaied},
  \citenamefont {Palacios}, \citenamefont {Salgado}, \citenamefont {Ugeda},
  \citenamefont {Veuillen}, \citenamefont {Yndurain},\ and\ \citenamefont
  {Brihuega}}]{Gonzalez2016}%
  \BibitemOpen
  \bibfield  {author} {\bibinfo {author} {\bibfnamefont {H.}~\bibnamefont
  {Gonz{\'a}lez-Herrero}}, \bibinfo {author} {\bibfnamefont {J.~M.}\
  \bibnamefont {G{\'o}mez-Rodr{\'\i}guez}}, \bibinfo {author} {\bibfnamefont
  {P.}~\bibnamefont {Mallet}}, \bibinfo {author} {\bibfnamefont
  {M.}~\bibnamefont {Moaied}}, \bibinfo {author} {\bibfnamefont {J.~J.}\
  \bibnamefont {Palacios}}, \bibinfo {author} {\bibfnamefont {C.}~\bibnamefont
  {Salgado}}, \bibinfo {author} {\bibfnamefont {M.~M.}\ \bibnamefont {Ugeda}},
  \bibinfo {author} {\bibfnamefont {J.-Y.}\ \bibnamefont {Veuillen}}, \bibinfo
  {author} {\bibfnamefont {F.}~\bibnamefont {Yndurain}}, \ and\ \bibinfo
  {author} {\bibfnamefont {I.}~\bibnamefont {Brihuega}},\ }\href {\doibase
  10.1126/science.aad8038} {\bibfield  {journal} {\bibinfo  {journal}
  {Science}\ }\textbf {\bibinfo {volume} {352}},\ \bibinfo {pages} {437}
  (\bibinfo {year} {2016})}\BibitemShut {NoStop}%
\bibitem [{\citenamefont {Sanchez-Valencia}\ \emph {et~al.}(2014)\citenamefont
  {Sanchez-Valencia}, \citenamefont {Dienel}, \citenamefont {Groning},
  \citenamefont {Shorubalko}, \citenamefont {Mueller}, \citenamefont {Jansen},
  \citenamefont {Amsharov}, \citenamefont {Ruffieux},\ and\ \citenamefont
  {Fasel}}]{SanchezValencia2014}%
  \BibitemOpen
  \bibfield  {author} {\bibinfo {author} {\bibfnamefont {J.~R.}\ \bibnamefont
  {Sanchez-Valencia}}, \bibinfo {author} {\bibfnamefont {T.}~\bibnamefont
  {Dienel}}, \bibinfo {author} {\bibfnamefont {O.}~\bibnamefont {Groning}},
  \bibinfo {author} {\bibfnamefont {I.}~\bibnamefont {Shorubalko}}, \bibinfo
  {author} {\bibfnamefont {A.}~\bibnamefont {Mueller}}, \bibinfo {author}
  {\bibfnamefont {M.}~\bibnamefont {Jansen}}, \bibinfo {author} {\bibfnamefont
  {K.}~\bibnamefont {Amsharov}}, \bibinfo {author} {\bibfnamefont
  {P.}~\bibnamefont {Ruffieux}}, \ and\ \bibinfo {author} {\bibfnamefont
  {R.}~\bibnamefont {Fasel}},\ }\href {http://dx.doi.org/10.1038/nature13607}
  {\bibfield  {journal} {\bibinfo  {journal} {Nature}\ }\textbf {\bibinfo
  {volume} {512}},\ \bibinfo {pages} {61} (\bibinfo {year} {2014})}\BibitemShut
  {NoStop}%
\bibitem [{\citenamefont {Ruffieux}\ \emph {et~al.}(2016)\citenamefont
  {Ruffieux}, \citenamefont {Wang}, \citenamefont {Yang}, \citenamefont
  {S{\'a}nchez-S{\'a}nchez}, \citenamefont {Liu}, \citenamefont {Dienel},
  \citenamefont {Talirz}, \citenamefont {Shinde}, \citenamefont {Pignedoli},
  \citenamefont {Passerone}, \citenamefont {Dumslaff}, \citenamefont {Feng},
  \citenamefont {M{\"u}llen},\ and\ \citenamefont {Fasel}}]{Ruffieux2016}%
  \BibitemOpen
  \bibfield  {author} {\bibinfo {author} {\bibfnamefont {P.}~\bibnamefont
  {Ruffieux}}, \bibinfo {author} {\bibfnamefont {S.}~\bibnamefont {Wang}},
  \bibinfo {author} {\bibfnamefont {B.}~\bibnamefont {Yang}}, \bibinfo {author}
  {\bibfnamefont {C.}~\bibnamefont {S{\'a}nchez-S{\'a}nchez}}, \bibinfo
  {author} {\bibfnamefont {J.}~\bibnamefont {Liu}}, \bibinfo {author}
  {\bibfnamefont {T.}~\bibnamefont {Dienel}}, \bibinfo {author} {\bibfnamefont
  {L.}~\bibnamefont {Talirz}}, \bibinfo {author} {\bibfnamefont
  {P.}~\bibnamefont {Shinde}}, \bibinfo {author} {\bibfnamefont {C.~A.}\
  \bibnamefont {Pignedoli}}, \bibinfo {author} {\bibfnamefont {D.}~\bibnamefont
  {Passerone}}, \bibinfo {author} {\bibfnamefont {T.}~\bibnamefont {Dumslaff}},
  \bibinfo {author} {\bibfnamefont {X.}~\bibnamefont {Feng}}, \bibinfo {author}
  {\bibfnamefont {K.}~\bibnamefont {M{\"u}llen}}, \ and\ \bibinfo {author}
  {\bibfnamefont {R.}~\bibnamefont {Fasel}},\ }\href
  {http://dx.doi.org/10.1038/nature17151} {\bibfield  {journal} {\bibinfo
  {journal} {Nature}\ }\textbf {\bibinfo {volume} {531}},\ \bibinfo {pages}
  {489} (\bibinfo {year} {2016})}\BibitemShut {NoStop}%
\bibitem [{\citenamefont {Robertson}\ \emph {et~al.}(2012)\citenamefont
  {Robertson}, \citenamefont {Allen}, \citenamefont {Wu}, \citenamefont {He},
  \citenamefont {Olivier}, \citenamefont {Neethling}, \citenamefont
  {Kirkland},\ and\ \citenamefont {Warner}}]{Robertson2012}%
  \BibitemOpen
  \bibfield  {author} {\bibinfo {author} {\bibfnamefont {A.~W.}\ \bibnamefont
  {Robertson}}, \bibinfo {author} {\bibfnamefont {C.~S.}\ \bibnamefont
  {Allen}}, \bibinfo {author} {\bibfnamefont {Y.~A.}\ \bibnamefont {Wu}},
  \bibinfo {author} {\bibfnamefont {K.}~\bibnamefont {He}}, \bibinfo {author}
  {\bibfnamefont {J.}~\bibnamefont {Olivier}}, \bibinfo {author} {\bibfnamefont
  {J.}~\bibnamefont {Neethling}}, \bibinfo {author} {\bibfnamefont {A.~I.}\
  \bibnamefont {Kirkland}}, \ and\ \bibinfo {author} {\bibfnamefont {J.~H.}\
  \bibnamefont {Warner}},\ }\href@noop {} {\bibfield  {journal} {\bibinfo
  {journal} {Nature Comm.}\ }\textbf {\bibinfo {volume} {3}},\ \bibinfo {pages}
  {1144} (\bibinfo {year} {2012})}\BibitemShut {NoStop}%
\bibitem [{\citenamefont {Chico}\ \emph {et~al.}(2015)\citenamefont {Chico},
  \citenamefont {Latg\'e},\ and\ \citenamefont {Brey}}]{Chico2015}%
  \BibitemOpen
  \bibfield  {author} {\bibinfo {author} {\bibfnamefont {L.}~\bibnamefont
  {Chico}}, \bibinfo {author} {\bibfnamefont {A.}~\bibnamefont {Latg\'e}}, \
  and\ \bibinfo {author} {\bibfnamefont {L.}~\bibnamefont {Brey}},\ }\href@noop
  {} {\bibfield  {journal} {\bibinfo  {journal} {Phys. Chem. Chem. Phys.}\
  }\textbf {\bibinfo {volume} {17}},\ \bibinfo {pages} {16469} (\bibinfo {year}
  {2015})}\BibitemShut {NoStop}%
\bibitem [{\citenamefont {Santos}\ \emph {et~al.}(2016)\citenamefont {Santos},
  \citenamefont {Latg\'e}, \citenamefont {Alvarellos},\ and\ \citenamefont
  {Chico}}]{HSantos2016}%
  \BibitemOpen
  \bibfield  {author} {\bibinfo {author} {\bibfnamefont {H.}~\bibnamefont
  {Santos}}, \bibinfo {author} {\bibfnamefont {A.}~\bibnamefont {Latg\'e}},
  \bibinfo {author} {\bibfnamefont {J.~E.}\ \bibnamefont {Alvarellos}}, \ and\
  \bibinfo {author} {\bibfnamefont {L.}~\bibnamefont {Chico}},\ }\href@noop {}
  {\bibfield  {journal} {\bibinfo  {journal} {Phys. Rev. B}\ }\textbf {\bibinfo
  {volume} {93}},\ \bibinfo {pages} {165424} (\bibinfo {year}
  {2016})}\BibitemShut {NoStop}%
\bibitem [{\citenamefont {Lieb}(1989)}]{Lieb1989}%
  \BibitemOpen
  \bibfield  {author} {\bibinfo {author} {\bibfnamefont {E.~H.}\ \bibnamefont
  {Lieb}},\ }\href@noop {} {\bibfield  {journal} {\bibinfo  {journal} {Phys.
  Rev. Lett.}\ }\textbf {\bibinfo {volume} {62}},\ \bibinfo {pages} {1201}
  (\bibinfo {year} {1989})}\BibitemShut {NoStop}%
\bibitem [{\citenamefont {Saremi}(2007)}]{Saremi2007}%
  \BibitemOpen
  \bibfield  {author} {\bibinfo {author} {\bibfnamefont {S.}~\bibnamefont
  {Saremi}},\ }\href@noop {} {\bibfield  {journal} {\bibinfo  {journal} {Phys.
  Rev. B}\ }\textbf {\bibinfo {volume} {76}},\ \bibinfo {pages} {184430}
  (\bibinfo {year} {2007})}\BibitemShut {NoStop}%
\bibitem [{\citenamefont {Palacios}\ \emph {et~al.}(2008)\citenamefont
  {Palacios}, \citenamefont {Fern\'andez-Rossier},\ and\ \citenamefont
  {Brey}}]{Palacios2008}%
  \BibitemOpen
  \bibfield  {author} {\bibinfo {author} {\bibfnamefont {J.~J.}\ \bibnamefont
  {Palacios}}, \bibinfo {author} {\bibfnamefont {J.}~\bibnamefont
  {Fern\'andez-Rossier}}, \ and\ \bibinfo {author} {\bibfnamefont
  {L.}~\bibnamefont {Brey}},\ }\href@noop {} {\bibfield  {journal} {\bibinfo
  {journal} {Phys. Rev. B}\ }\textbf {\bibinfo {volume} {77}},\ \bibinfo
  {pages} {195428} (\bibinfo {year} {2008})}\BibitemShut {NoStop}%
\bibitem [{\citenamefont {Santos}\ \emph {et~al.}(2014)\citenamefont {Santos},
  \citenamefont {Soriano},\ and\ \citenamefont {Palacios}}]{HSantos2014}%
  \BibitemOpen
  \bibfield  {author} {\bibinfo {author} {\bibfnamefont {H.}~\bibnamefont
  {Santos}}, \bibinfo {author} {\bibfnamefont {D.}~\bibnamefont {Soriano}}, \
  and\ \bibinfo {author} {\bibfnamefont {J.~J.}\ \bibnamefont {Palacios}},\
  }\href@noop {} {\bibfield  {journal} {\bibinfo  {journal} {Phys. Rev. B}\
  }\textbf {\bibinfo {volume} {89}},\ \bibinfo {pages} {195416} (\bibinfo
  {year} {2014})}\BibitemShut {NoStop}%
\bibitem [{\citenamefont {Liang}\ and\ \citenamefont {Sofo}(2012)}]{Liang2012}%
  \BibitemOpen
  \bibfield  {author} {\bibinfo {author} {\bibfnamefont {S.-Z.}\ \bibnamefont
  {Liang}}\ and\ \bibinfo {author} {\bibfnamefont {J.~O.}\ \bibnamefont
  {Sofo}},\ }\href {\doibase 10.1103/PhysRevLett.109.256601} {\bibfield
  {journal} {\bibinfo  {journal} {Phys. Rev. Lett.}\ }\textbf {\bibinfo
  {volume} {109}},\ \bibinfo {pages} {256601} (\bibinfo {year}
  {2012})}\BibitemShut {NoStop}%
\bibitem [{\citenamefont {Hueso}\ \emph {et~al.}(2007)\citenamefont {Hueso},
  \citenamefont {Pruneda}, \citenamefont {Ferrari}, \citenamefont {Burnell},
  \citenamefont {Vald\'es-Herrera}, \citenamefont {Simmons}, \citenamefont
  {Littlewood}, \citenamefont {Artacho}, \citenamefont {Fert},\ and\
  \citenamefont {Mathur}}]{Hueso2007}%
  \BibitemOpen
  \bibfield  {author} {\bibinfo {author} {\bibfnamefont {L.}~\bibnamefont
  {Hueso}}, \bibinfo {author} {\bibfnamefont {J.~M.}\ \bibnamefont {Pruneda}},
  \bibinfo {author} {\bibfnamefont {V.}~\bibnamefont {Ferrari}}, \bibinfo
  {author} {\bibfnamefont {G.}~\bibnamefont {Burnell}}, \bibinfo {author}
  {\bibfnamefont {J.~P.}\ \bibnamefont {Vald\'es-Herrera}}, \bibinfo {author}
  {\bibfnamefont {B.~D.}\ \bibnamefont {Simmons}}, \bibinfo {author}
  {\bibfnamefont {P.~B.}\ \bibnamefont {Littlewood}}, \bibinfo {author}
  {\bibfnamefont {E.}~\bibnamefont {Artacho}}, \bibinfo {author} {\bibfnamefont
  {A.}~\bibnamefont {Fert}}, \ and\ \bibinfo {author} {\bibfnamefont {N.~D.}\
  \bibnamefont {Mathur}},\ }\href@noop {} {\bibfield  {journal} {\bibinfo
  {journal} {Nature (London)}\ }\textbf {\bibinfo {volume} {445}},\ \bibinfo
  {pages} {410} (\bibinfo {year} {2007})}\BibitemShut {NoStop}%
\bibitem [{\citenamefont {Saito}\ \emph {et~al.}(1992)\citenamefont {Saito},
  \citenamefont {Fujita}, \citenamefont {Dresselhaus},\ and\ \citenamefont
  {Dresselhaus}}]{Saito1992}%
  \BibitemOpen
  \bibfield  {author} {\bibinfo {author} {\bibfnamefont {R.}~\bibnamefont
  {Saito}}, \bibinfo {author} {\bibfnamefont {M.}~\bibnamefont {Fujita}},
  \bibinfo {author} {\bibfnamefont {G.}~\bibnamefont {Dresselhaus}}, \ and\
  \bibinfo {author} {\bibfnamefont {M.}~\bibnamefont {Dresselhaus}},\ }\href
  {\doibase 10.1063/1.107080} {\bibfield  {journal} {\bibinfo  {journal} {Appl.
  Phys. Lett.}\ }\textbf {\bibinfo {volume} {60}},\ \bibinfo {pages} {2204}
  (\bibinfo {year} {1992})}\BibitemShut {NoStop}%
\bibitem [{\citenamefont {Chico}\ \emph {et~al.}(1996)\citenamefont {Chico},
  \citenamefont {Crespi}, \citenamefont {Benedict}, \citenamefont {Louie},\
  and\ \citenamefont {Cohen}}]{Chico1996a}%
  \BibitemOpen
  \bibfield  {author} {\bibinfo {author} {\bibfnamefont {L.}~\bibnamefont
  {Chico}}, \bibinfo {author} {\bibfnamefont {V.~H.}\ \bibnamefont {Crespi}},
  \bibinfo {author} {\bibfnamefont {L.~X.}\ \bibnamefont {Benedict}}, \bibinfo
  {author} {\bibfnamefont {S.~G.}\ \bibnamefont {Louie}}, \ and\ \bibinfo
  {author} {\bibfnamefont {M.~L.}\ \bibnamefont {Cohen}},\ }\href@noop {}
  {\bibfield  {journal} {\bibinfo  {journal} {Phys. Rev. Lett.}\ }\textbf
  {\bibinfo {volume} {76}},\ \bibinfo {pages} {971} (\bibinfo {year}
  {1996})}\BibitemShut {NoStop}%
\bibitem [{\citenamefont {Charlier}\ \emph {et~al.}(1996)\citenamefont
  {Charlier}, \citenamefont {Ebbesen},\ and\ \citenamefont
  {Lambin}}]{Charlier1996}%
  \BibitemOpen
  \bibfield  {author} {\bibinfo {author} {\bibfnamefont {J.-C.}\ \bibnamefont
  {Charlier}}, \bibinfo {author} {\bibfnamefont {T.~W.}\ \bibnamefont
  {Ebbesen}}, \ and\ \bibinfo {author} {\bibfnamefont {P.}~\bibnamefont
  {Lambin}},\ }\href {\doibase 10.1103/PhysRevB.53.11108} {\bibfield  {journal}
  {\bibinfo  {journal} {Phys. Rev. B}\ }\textbf {\bibinfo {volume} {53}},\
  \bibinfo {pages} {11108} (\bibinfo {year} {1996})}\BibitemShut {NoStop}%
\bibitem [{\citenamefont {Qiao}\ \emph {et~al.}(2010)\citenamefont {Qiao},
  \citenamefont {Yang}, \citenamefont {Feng}, \citenamefont {Tse},
  \citenamefont {Ding}, \citenamefont {Yao}, \citenamefont {Wang},\ and\
  \citenamefont {Niu}}]{Qiao2010}%
  \BibitemOpen
  \bibfield  {author} {\bibinfo {author} {\bibfnamefont {Z.}~\bibnamefont
  {Qiao}}, \bibinfo {author} {\bibfnamefont {S.}~\bibnamefont {Yang}}, \bibinfo
  {author} {\bibfnamefont {W.}~\bibnamefont {Feng}}, \bibinfo {author}
  {\bibfnamefont {W.-K.}\ \bibnamefont {Tse}}, \bibinfo {author} {\bibfnamefont
  {J.}~\bibnamefont {Ding}}, \bibinfo {author} {\bibfnamefont {Y.}~\bibnamefont
  {Yao}}, \bibinfo {author} {\bibfnamefont {J.}~\bibnamefont {Wang}}, \ and\
  \bibinfo {author} {\bibfnamefont {Q.}~\bibnamefont {Niu}},\ }\href {\doibase
  10.1103/PhysRevB.82.161414} {\bibfield  {journal} {\bibinfo  {journal} {Phys.
  Rev. B}\ }\textbf {\bibinfo {volume} {82}},\ \bibinfo {pages} {161414}
  (\bibinfo {year} {2010})}\BibitemShut {NoStop}%
\bibitem [{\citenamefont {Lenz}\ \emph {et~al.}(2013)\citenamefont {Lenz},
  \citenamefont {Urban},\ and\ \citenamefont {Bercioux}}]{Lenz2013}%
  \BibitemOpen
  \bibfield  {author} {\bibinfo {author} {\bibfnamefont {L.}~\bibnamefont
  {Lenz}}, \bibinfo {author} {\bibfnamefont {D.~F.}\ \bibnamefont {Urban}}, \
  and\ \bibinfo {author} {\bibfnamefont {D.}~\bibnamefont {Bercioux}},\ }\href
  {http://dx.doi.org/10.1140/epjb/e2013-40760-4} {\bibfield  {journal}
  {\bibinfo  {journal} {Eur. Phys. J. B}\ }\textbf {\bibinfo {volume} {86}},\
  \bibinfo {pages} {502} (\bibinfo {year} {2013})}\BibitemShut {NoStop}%
\bibitem [{\citenamefont {Marchenko}\ \emph {et~al.}(2012)\citenamefont
  {Marchenko}, \citenamefont {Varykhalov}, \citenamefont {Scholz},
  \citenamefont {Bihlmayer}, \citenamefont {Rashba}, \citenamefont {Rybkin},
  \citenamefont {Shikin},\ and\ \citenamefont {Rader}}]{Marchenko2012}%
  \BibitemOpen
  \bibfield  {author} {\bibinfo {author} {\bibfnamefont {D.}~\bibnamefont
  {Marchenko}}, \bibinfo {author} {\bibfnamefont {A.}~\bibnamefont
  {Varykhalov}}, \bibinfo {author} {\bibfnamefont {M.~R.}\ \bibnamefont
  {Scholz}}, \bibinfo {author} {\bibfnamefont {G.}~\bibnamefont {Bihlmayer}},
  \bibinfo {author} {\bibfnamefont {E.~I.}\ \bibnamefont {Rashba}}, \bibinfo
  {author} {\bibfnamefont {A.}~\bibnamefont {Rybkin}}, \bibinfo {author}
  {\bibfnamefont {A.~M.}\ \bibnamefont {Shikin}}, \ and\ \bibinfo {author}
  {\bibfnamefont {O.}~\bibnamefont {Rader}},\ }\href
  {http://dx.doi.org/10.1038/ncomms2227} {\bibfield  {journal} {\bibinfo
  {journal} {Nat. Commun.}\ }\textbf {\bibinfo {volume} {3}},\ \bibinfo {pages}
  {1232} (\bibinfo {year} {2012})}\BibitemShut {NoStop}%
\bibitem [{\citenamefont {Balakrishnan}\ \emph {et~al.}(2014)\citenamefont
  {Balakrishnan}, \citenamefont {Koon}, \citenamefont {Avsar}, \citenamefont
  {Ho}, \citenamefont {Lee}, \citenamefont {Jaiswal}, \citenamefont {Baeck},
  \citenamefont {Ahn}, \citenamefont {Ferreira}, \citenamefont {Cazalilla},
  \citenamefont {Castro~Neto},\ and\ \citenamefont
  {{\"O}zyilmaz}}]{Balakrishnan2014}%
  \BibitemOpen
  \bibfield  {author} {\bibinfo {author} {\bibfnamefont {J.}~\bibnamefont
  {Balakrishnan}}, \bibinfo {author} {\bibfnamefont {G.~K.~W.}\ \bibnamefont
  {Koon}}, \bibinfo {author} {\bibfnamefont {A.}~\bibnamefont {Avsar}},
  \bibinfo {author} {\bibfnamefont {Y.}~\bibnamefont {Ho}}, \bibinfo {author}
  {\bibfnamefont {J.~H.}\ \bibnamefont {Lee}}, \bibinfo {author} {\bibfnamefont
  {M.}~\bibnamefont {Jaiswal}}, \bibinfo {author} {\bibfnamefont {S.-J.}\
  \bibnamefont {Baeck}}, \bibinfo {author} {\bibfnamefont {J.-H.}\ \bibnamefont
  {Ahn}}, \bibinfo {author} {\bibfnamefont {A.}~\bibnamefont {Ferreira}},
  \bibinfo {author} {\bibfnamefont {M.~A.}\ \bibnamefont {Cazalilla}}, \bibinfo
  {author} {\bibfnamefont {A.~H.}\ \bibnamefont {Castro~Neto}}, \ and\ \bibinfo
  {author} {\bibfnamefont {B.}~\bibnamefont {{\"O}zyilmaz}},\ }\href
  {http://dx.doi.org/10.1038/ncomms5748} {\bibfield  {journal} {\bibinfo
  {journal} {Nat. Commun.}\ }\textbf {\bibinfo {volume} {5}},\ \bibinfo {pages}
  {4748} (\bibinfo {year} {2014})}\BibitemShut {NoStop}%
\bibitem [{\citenamefont {R.Winkler}(2003)}]{WinklerBook}%
  \BibitemOpen
  \bibfield  {author} {\bibinfo {author} {\bibnamefont {R.Winkler}},\
  }\href@noop {} {\emph {\bibinfo {title} {Spin-Orbit Coupling Effects in
  Two-Dimensional Electron and Hole Systems}}}\ (\bibinfo  {publisher}
  {Springer-Verlag Berlin},\ \bibinfo {year} {2003})\BibitemShut {NoStop}%
\bibitem [{\citenamefont {Xu}\ \emph {et~al.}(2007)\citenamefont {Xu},
  \citenamefont {Li}, \citenamefont {Pan},\ and\ \citenamefont {Zhu}}]{Xu2007}%
  \BibitemOpen
  \bibfield  {author} {\bibinfo {author} {\bibfnamefont {F.}~\bibnamefont
  {Xu}}, \bibinfo {author} {\bibfnamefont {B.}~\bibnamefont {Li}}, \bibinfo
  {author} {\bibfnamefont {H.}~\bibnamefont {Pan}}, \ and\ \bibinfo {author}
  {\bibfnamefont {J.-L.}\ \bibnamefont {Zhu}},\ }\href {\doibase
  10.1103/PhysRevB.75.085431} {\bibfield  {journal} {\bibinfo  {journal} {Phys.
  Rev. B}\ }\textbf {\bibinfo {volume} {75}},\ \bibinfo {pages} {085431}
  (\bibinfo {year} {2007})}\BibitemShut {NoStop}%
\bibitem [{\citenamefont {Diniz}\ \emph {et~al.}(2012)\citenamefont {Diniz},
  \citenamefont {Latg\'e},\ and\ \citenamefont {Ulloa}}]{Diniz2012}%
  \BibitemOpen
  \bibfield  {author} {\bibinfo {author} {\bibfnamefont {G.~S.}\ \bibnamefont
  {Diniz}}, \bibinfo {author} {\bibfnamefont {A.}~\bibnamefont {Latg\'e}}, \
  and\ \bibinfo {author} {\bibfnamefont {S.~E.}\ \bibnamefont {Ulloa}},\ }\href
  {\doibase 10.1103/PhysRevLett.108.126601} {\bibfield  {journal} {\bibinfo
  {journal} {Phys. Rev. Lett.}\ }\textbf {\bibinfo {volume} {108}},\ \bibinfo
  {pages} {126601} (\bibinfo {year} {2012})}\BibitemShut {NoStop}%
\bibitem [{\citenamefont {Zhai}\ and\ \citenamefont {Xu}(2005)}]{Zhai2005}%
  \BibitemOpen
  \bibfield  {author} {\bibinfo {author} {\bibfnamefont {F.}~\bibnamefont
  {Zhai}}\ and\ \bibinfo {author} {\bibfnamefont {H.~Q.}\ \bibnamefont {Xu}},\
  }\href@noop {} {\bibfield  {journal} {\bibinfo  {journal} {Phys. Rev. Lett.}\
  }\textbf {\bibinfo {volume} {94}},\ \bibinfo {pages} {246601} (\bibinfo
  {year} {2005})}\BibitemShut {NoStop}%
\bibitem [{\citenamefont {Paul L.~McEuen}\ \emph {et~al.}(2016)\citenamefont
  {Paul L.~McEuen}, \citenamefont {Bockrath}, \citenamefont {Yoon},\ and\
  \citenamefont {Louie}}]{Paul1999}%
  \BibitemOpen
  \bibfield  {author} {\bibinfo {author} {\bibfnamefont {P.~L.}\ \bibnamefont
  {Paul L.~McEuen}}, \bibinfo {author} {\bibfnamefont {D.~H.}\ \bibnamefont
  {Bockrath}, \bibfnamefont {Marc and.~Cobden}}, \bibinfo {author}
  {\bibfnamefont {Y.-G.}\ \bibnamefont {Yoon}}, \ and\ \bibinfo {author}
  {\bibfnamefont {S.~G.}\ \bibnamefont {Louie}},\ }\href {\doibase
  /10.1103/PhysRevLett.83.5098} {\bibfield  {journal} {\bibinfo  {journal}
  {Phys. Rev. Lett.}\ }\textbf {\bibinfo {volume} {83}},\ \bibinfo {pages}
  {5098} (\bibinfo {year} {2016})}\BibitemShut {NoStop}%
\bibitem [{\citenamefont {Garc\'{\i}a-Lastra}\ \emph
  {et~al.}(2008)\citenamefont {Garc\'{\i}a-Lastra}, \citenamefont {Thygesen},
  \citenamefont {Strange},\ and\ \citenamefont {Rubio}}]{GarciaLastra2008}%
  \BibitemOpen
  \bibfield  {author} {\bibinfo {author} {\bibfnamefont {J.~M.}\ \bibnamefont
  {Garc\'{\i}a-Lastra}}, \bibinfo {author} {\bibfnamefont {K.~S.}\ \bibnamefont
  {Thygesen}}, \bibinfo {author} {\bibfnamefont {M.}~\bibnamefont {Strange}}, \
  and\ \bibinfo {author} {\bibfnamefont {A.}~\bibnamefont {Rubio}},\
  }\href@noop {} {\bibfield  {journal} {\bibinfo  {journal} {Phys. Rev. Lett.}\
  }\textbf {\bibinfo {volume} {101}},\ \bibinfo {pages} {236806} (\bibinfo
  {year} {2008})}\BibitemShut {NoStop}%
\bibitem [{\citenamefont {Khalfoun}\ \emph {et~al.}(2014)\citenamefont
  {Khalfoun}, \citenamefont {Lambin},\ and\ \citenamefont
  {Henrard}}]{Khalfoun2014}%
  \BibitemOpen
  \bibfield  {author} {\bibinfo {author} {\bibfnamefont {H.}~\bibnamefont
  {Khalfoun}}, \bibinfo {author} {\bibfnamefont {P.}~\bibnamefont {Lambin}}, \
  and\ \bibinfo {author} {\bibfnamefont {L.}~\bibnamefont {Henrard}},\
  }\href@noop {} {\bibfield  {journal} {\bibinfo  {journal} {Phys. Rev. B}\
  }\textbf {\bibinfo {volume} {89}},\ \bibinfo {pages} {045407} (\bibinfo
  {year} {2014})}\BibitemShut {NoStop}%
\bibitem [{\citenamefont {S.Datta}(1997)}]{DattaBook}%
  \BibitemOpen
  \bibfield  {author} {\bibinfo {author} {\bibnamefont {S.Datta}},\ }\href@noop
  {} {\emph {\bibinfo {title} {Electronic Transport in Mesoscopic Systems}}}\
  (\bibinfo  {publisher} {Cambridge University Press},\ \bibinfo {year}
  {1997})\BibitemShut {NoStop}%
\bibitem [{\citenamefont {Santos}\ \emph {et~al.}(2009)\citenamefont {Santos},
  \citenamefont {Chico},\ and\ \citenamefont {Brey}}]{SCB09}%
  \BibitemOpen
  \bibfield  {author} {\bibinfo {author} {\bibfnamefont {H.}~\bibnamefont
  {Santos}}, \bibinfo {author} {\bibfnamefont {L.}~\bibnamefont {Chico}}, \
  and\ \bibinfo {author} {\bibfnamefont {L.}~\bibnamefont {Brey}},\ }\href
  {\doibase 10.1103/PhysRevLett.103.086801} {\bibfield  {journal} {\bibinfo
  {journal} {Phys. Rev. Lett.}\ }\textbf {\bibinfo {volume} {103}},\ \bibinfo
  {pages} {086801} (\bibinfo {year} {2009})}\BibitemShut {NoStop}%
\bibitem [{\citenamefont {Alam}\ and\ \citenamefont
  {Pramanik}(2015)}]{Alam2015}%
  \BibitemOpen
  \bibfield  {author} {\bibinfo {author} {\bibfnamefont {K.~M.}\ \bibnamefont
  {Alam}}\ and\ \bibinfo {author} {\bibfnamefont {S.}~\bibnamefont
  {Pramanik}},\ }\href@noop {} {\bibfield  {journal} {\bibinfo  {journal} {Adv.
  Funct. Mater.}\ }\textbf {\bibinfo {volume} {25}},\ \bibinfo {pages} {3210}
  (\bibinfo {year} {2015})}\BibitemShut {NoStop}%
\end{thebibliography}

%

\end{document}